\documentclass[traditabstract]{aa}
\usepackage{graphicx}
\usepackage{txfonts}
\usepackage{longtable}
\usepackage{xcolor}
\usepackage{natbib}
\usepackage{threeparttable}
\usepackage{pdfpages}
\usepackage{verbatim}

\bibpunct{(}{)}{;}{a}{}{,}

\begin{document} 

\newcommand{\hi}{\mbox{H\,{\sc i}}}
\newcommand{\zabs}{$z_{\rm abs}$}
\newcommand{\zmin}{$z_{\rm min}$}
\newcommand{\zmax}{$z_{\rm max}$}
\newcommand{\zq}{$z_{\rm q}$}
\newcommand{\zg}{$z_{\rm g}$}
\newcommand{\kms}{km\,s$^{-1}$}
\newcommand{\cmsq}{cm$^{-2}$}
\newcommand{\degree}{\ensuremath{^\circ}}
\newcommand{\Msun}{$M_{\odot}$} 
\newcommand{\mgii}{\mbox{Mg\,{\sc ii}}} 
\newcommand{\mgiia}{\mbox{Mg\,{\sc ii}$\lambda$2796}}
\newcommand{\mgiib}{\mbox{Mg\,{\sc ii}$\lambda$2803}}
\newcommand{\mgiiab}{\mbox{Mg\,{\sc ii}$\lambda\lambda$2796,2803}}
\newcommand{\lapp}{\mbox{\raisebox{-0.3em}{$\stackrel{\textstyle <}{\sim}$}}}
\newcommand{\gapp}{\mbox{\raisebox{-0.3em}{$\stackrel{\textstyle >}{\sim}$}}}
\newcommand{\pks}{PKS\,1200$+$045}
\newcommand{\pkss}{PKS\,1245$-$19}
\def\hh     {\ifmmode{{\rm H}_2}\else{H$_2$}\fi}

\newcommand{\lerma}{Observatoire de Paris, LERMA, Coll\`ege de France, CNRS, PSL University, Sorbonne University, 75014, Paris --- \email{francoise.combes@obspm.fr} \label{lerma}}
\newcommand{\iucaa}{Inter-University Centre for Astronomy and Astrophysics, Post Bag 4, Ganeshkhind, Pune 411 007, India \label{iucaa}}

\definecolor{green}{rgb}{0,0.4,0}

\titlerunning{Cold molecules in \hi\ absorbers across $z$ = 0.1 - 4}
\authorrunning{F. Combes, N. Gupta}

   \title{Cold molecules in \hi\ 21cm absorbers across redshifts $\sim$ 0.1 - 4
   \thanks{Based on observations carried out with the IRAM-30m telescope and NOEMA, the NOrthern Extended Millimeter Array -- IRAM (Institute of RAdioastronomy in Millimeter).  }
   }
   \author{
        F. Combes\inst{\ref{lerma}}
            \and
        N. Gupta\inst{\ref{iucaa}}
           }

    \institute{\lerma \and \iucaa 
    }

   \date{Received: October 2023; accepted: December 2023}
 
  \abstract {
Absorption lines at high redshift in front of quasars are quite rare 
in the millimeter (mm) domain.  Only five associated and five intervening 
systems have been reported in the literature.
Nevertheless, these discoveries provide very useful information that is
complementary to emission lines, allowing, for instance, to distinguish
between inflows and outflows. These lines are also good candidates for studying the 
variations of the fundamental constants of physics. 
Here we report the findings of our search for CO and other molecules
in emission and absorption in front of a sample of 30 targets, comprising
16 associated and 14 intervening \hi\ 21cm absorbers. 
 The observations were made with the IRAM-30m telescope
 simultaneously at 3mm and 2mm, exploring several lines of the CO ladder 
 and HCO$^+$, depending on the redshift. We detected eight targets 
 in emission, of which five are new.  The derived molecular gas masses 
 range from 10$^9$ to 7$\times$ 10$^{11}$ M$_\odot$ and the highest
 redshift detection ($z$ = 3.387) corresponds to a relatively 
 average-metallicity damped Lyman-$\alpha$ absorber for this redshift.    
 We also report four new detections in absorption.
 Two of the associated CO absorption line detections at high redshift ($z$=1.211 and 1.275)
 result from high-spatial-resolution follow-up observations with NOEMA. 
 The disparity between the mm molecular and \hi\ 21cm absorption lines for these and another intervening system detected in HNC at $z=1.275$ is attributable to radio and mm sight lines tracing different media.  
We compare the atomic and molecular column densities of 14 known high-redshift ($z$ > 0.1)
molecular absorption line systems. The associated \hi\ absorption lines are broad and exhibit multiple components, and the molecular absorption generally corresponds to the broader 
and weaker 21cm absorption component. This indicates two distinct phases: one near galaxy
centers with a larger CO-to-\hi\ abundance ratio, and another with lower molecular
abundance in the outer regions of the galaxy.  
In comparison, intervening absorption profiles correspond primarily to \hi-dominated gas structure 
in galaxy outskirts, except for gas at low impact parameters in gravitationally lensed systems.
The comparison of interferometric and single-dish observations presented here 
shows that the detection of absorption requires sufficient 
 spatial resolution to overcome the dilution by emission and will be an important criterion for mm follow-up of 21cm absorbers from ongoing large-scale surveys.
}
   

\maketitle
%
\section{Introduction}
\label{sec:intro}

The star formation history of the Universe is characterized by
a peak about 10 billion years ago. The star formation rate (SFR) density
in galaxies, as reviewed by \citet{Madau2014}, evolves across cosmic time:
it first increases with time ($z>$2), peaks at $z\sim$1-2, and then
decreases by an order of magnitude from  z$\sim$1 to the present day.
Two main factors are suspected to produce this evolution: the gas
content of galaxies, and their star formation efficiency (SFE).
 From studies of main sequence star-forming galaxies across the Hubble time,
 the gas fraction in the interstellar medium (ISM) of galaxies appears to be the dominant factor in comparison to the SFE
 \citep[e.g.][]{Tacconi2018}. It is therefore crucial to constrain the
molecular gas content of galaxies over cosmic time, 
and also its precursor, that is,
the atomic gas reservoir. \citet{Obreschkow2009a,Obreschkow2009b}
used theoretical models and the knowledge of molecular gas content
of high-redshift galaxies to make very useful predictions of the cosmic evolution 
of \hi\ and H$_2$ gas components. The \hi\ evolution was found to be relatively
constant over $z$=1-5 from damped Lyman-$\alpha$ absorber (DLA) studies, while the H$_2$/HI ratio
was steeply declining with time, as (1+$z$)$^{1.6}$. Since then, many molecular line surveys
have refined our knowledge, and shown that the cosmic H$_2$ evolution
might be flatter \citep{Decarli2020,Riechers2020, Lenkic2020}. 
In this context, it is important to pursue blind surveys, such as serendipitous discoveries
of CO lines in the field of view of the Atacama Large Millimeter/submillimeter Array (ALMA) calibration sources \citep[e.g.,][]{Hamanowicz2023}, to obtain an unbiased view of the gas evolution.

Our knowledge of molecular gas in galaxies at high redshift has received major boosts since the first 
discovery by \citet{Solomon1992}, first by gravitational lensing, 
as reviewed by \citet{Solomon2005}, and then through the negative K-correction thanks to which 
the CO molecule can be observed over the entirety of its rotation ladder, 
with the flux density increasing strongly with the J-level \citep[e.g.,][]{Combes1999}. 
Although more difficult to detect, molecular absorption lines can bring new and complementary information. 
While the strength of emission lines decreases as a steep function of redshift, absorption lines can be observed as easily at high redshift as in the local Universe provided there exists a strong background radio source \citep[e.g.,][]{Combes2008}. Further, while emission lines are sensitive to dense and warm molecular gas, absorption lines arise in the low-excitation and diffuse gas \citep[e.g.,][]{Wiklind1994, Wiklind1996nat, Wiklind1997, Menten2008, Henkel2009, Muller2014, Wiklind2018}.  In conclusion, with a 3mm continuum source flux density of at least $\sim$ 50~mJy, absorption lines can be potentially used to obtain information about molecular gas in galaxies undetectable through emission lines.

Absorption lines also confer the advantage of the ease of follow-up observations in many molecular lines other than CO. Indeed, once a high-column-density absorber has been discovered, the detection of other lines does not suffer from low filling factor ---as is the case for emission---, nor critical density for excitation. This enables characterization of the physical and chemical conditions in the absorbing gas \citep[e.g.][]{Henkel2005, Bottinelli2009, Muller2014, Muller2016,  Muller2021}. The relative strengths of species like H$_3$CN, where the excitation is dominated by the cosmic microwave background (CMB), can be used to determine the CMB temperature \citep[e.g.][]{Henkel2009, Muller2013}.
Comparisons between the redshifts of different transitions (e.g. NH$_3$, CH$_3$OH, OH, etc.) can be used to test for variations in the fundamental
constants \citep[e.g.][]{Uzan2011, Kanekar2011, Kanekar2012, Bagdonaite2013, Murphy2022}.

Absorption-line studies broadly fall in two categories.  In cases where the absorbing gas ($z_{abs}$) is located in the same radio source ($z_{em}$), i.e., an active galactic nucleus (AGN), the absorber is called an {associated} absorber ($z_{abs}$ $\sim$ $z_{em}$). 
 According to the adopted definition, the relative difference between $z_{em}$ and $z_{abs}$ may only be lower than 3000 \kms \citep[e.g.,][]{Wolfe1986, Ellison2002, Gupta2021}.
In this category, the absorbing gas may originate from the circumnuclear disk or may represent outflowing or infalling material.
In the second category, the absorbing gas is unrelated to the background AGN and arises from an {``intervening''} galaxy ($z_{abs}$ $<$ $z_{em}$).
Through high-spatial-resolution observations, the number of associated molecular absorption line detections in the local Universe is steadily increasing \citep[e.g.][]{Tremblay2016, Rose2019Hydra, Rose2019all, Rose2020}. These provide extremely useful information about the gas distribution and kinematics, which can be used, for example, to distinguish inflows from outflows. Two absorbers are also detected  at high redshift: Abell 2390 ($z$=0.230) and RXCJ0439.0+0520 \citep[$z$=0.208; ][]{Rose2019all}; both  are associated with the brightest galaxies of the corresponding galaxy cluster.

Molecular absorptions at higher redshifts are extremely rare.
Up to now, only ten molecular absorption systems have been discovered  in the mm domain
at $z$$\gtrsim$0.1. Among these, five are intervening molecular absorptions: three are associated with lensing galaxies
amplifying the background quasars PKS1830-211, B0218+357 \citep[e.g.][]{Combes2008},
and PMN0134-0931 \citep{Wiklind2018}; the fourth is PKS1413+135 and the last is our recent CO/CN absorption detection
in front of the quasar Q0248+430 \citep[$z_{em}$=1.31;][]{Combes2019}.  The quasar sight line in the latter case is probing the gas associated with
the tidal tail emanating from a merging galaxy pair, G0248+430 ($z$ = 0.05194), at an impact parameter of $\sim$15\,kpc.
The \hi\ and OH absorptions at centimeter wavelengths have already been detected, revealing the presence of cold atomic and molecular gas toward the sight line \citep{Gupta2018oh}, leading to the above-mentioned CO/CN detection with the the NOrthern Extended Millimeter Array (NOEMA).

The five high-$z$ associated absorptions, in addition to the two associated with the brightest cluster galaxies (BCGs) mentioned above, are B3 1504+377 \citep{Combes2008}, 
4C+12.50 \citep{Dasyra2012}, and PKS B1740-517 \citep{Allison2019}. 
This recent CO detection with the ALMA was motivated by a \hi\ 21cm absorption detection \citep{Allison2015}. 
The CO(3-2) spectrum of 4C+12.50 displays both an emission component at the galaxy redshift $z$ = 0.1218 and an absorption feature of $\sim$ 1000\,\kms\ from it.  
The latter was interpreted by \citet{Dasyra2012}
as representing outflowing gas. Recently, a search for associated absorption lines was
also attempted with the calibrators of ALMA, but with no success \citep[ALMACAL;][]{Klitsch2019}. We also note that it is debatable whether the absorber toward PKS1413+135 is of associated or intervening category \citep{Readhead2021,Combes2023}. Based on recent advances, here we consider it to be an intervening absorber.

Interestingly, molecular absorptions of H$_2$ and CO are relatively easy to detect at high redshift using ultraviolet (UV) lines \citep[e.g.][]{Petitjean2000, Srianand2008, Noterdaeme2023}. 
The H$_2$ detection rates among sight lines selected on the basis of the presence of a DLA ($N$(\hi)$>2\times10^{20}$\,\cmsq) are low ($\lesssim$10\%). H$_2$ is more frequently detected in DLAs with higher metallicity.  More importantly, UV lines can detect molecular column densities one or two orders of magnitude lower ($N$(H$_2$)$\sim10^{14}$\,\cmsq) than what is possible with mm lines.  But at higher column densities, the background source is obscured by accompanying dust \citep{Noterdaeme2015}. As UV lines are biased against high-column-density molecular gas, the mm absorption lines are a welcome complementary tool \citep{Combes2008}.

We report here a systematic search for CO emission and absorption in a sample of 30 targets selected on the basis of the presence of cold atomic gas already confirmed through the detection of \hi\ 21cm absorption.  
This paper is structured as follows. Our sample is described in Sect. \ref{sec:samp}. Section \ref{sec:obs} presents the details of our mm observations using the Institut de Radioastronomie Millimetrique (IRAM)-30m telescope and NOEMA. The results in terms of optical depth and column density are quantified in Section \ref{sec:res}. In
Sect. \ref{sec:disc}, we discuss the constraints brought by the negative results on the various absorption components in comparison with the atomic or ionized gas absorptions. Section \ref{conclu} summarizes our conclusions.
Throughout the present paper, the velocity scale is defined with respect to redshifts indicated in Table
\ref{tab:samp}.  To compute distances, we adopt a flat $\Lambda$CDM cosmology,
with $\Omega_m$=0.29, $\Omega_\Lambda$=0.71, and the Hubble constant
H$_0$ = 70~\kms Mpc$^{-1}$. 

\section{The sample}
\label{sec:samp}

We gathered a sample of 30 candidates for molecular absorption and emission
line search.  In all of them, the presence of cold atomic gas has already been confirmed
through detection of \hi\ 21cm absorption (Table \ref{tab:samp}). Overall, the sample consists of 16 associated systems
at the redshift z$_{abs}\sim$ z$_{em}$, and 14 intervening ones
where z$_{abs}$ is significantly lower than z$_{em}$. 
The selected AGNs typically have high ($>$ 100 mJy) flux densities at cm wavelengths.
The exceptions are J0229+0044 (3.4\,mJy) and J0229+0053 (31.1\,mJy) from \citet[][]{Chowdhury2020} in which \hi\ 21cm absorption is still detected due to the presence of large amounts of cold gas.
At low-$z$ ($z \sim$ 0.1), such high 21cm optical depths are found to be associated with radio sources embedded in merging
galaxy pairs or Ultra Luminous Infrared Galaxies \citep[ULIRGs;][]{Dutta2019}.  The basic properties
of the sample are presented in Table \ref{tab:samp}.
 The absorption redshifts, $z_{abs}$, in Table~\ref{tab:samp} are based on the peak of \hi\ 21cm absorption lines, except for \pks, where it is the middle of the absorption range; this line is
 highly blueshifted with respect to the emission redshift. The intervening absorption lines are generally narrow with widths of $\sim$50\,\kms\  and these are therefore observed with spectral resolution of a few \kms\ \citep[e.g.,][]{Gupta2012}.  The absorption redshifts of these absorbers in Table~\ref{tab:samp} have higher accuracy compared to the associated absorbers, which are relatively broader ($\sim$100\,\kms) and observed with coarser spectral resolution. The interpretation of gas detected in associated absorption crucially depends on the measurement of systemic redshift, $z_{em}$, based on optical emission lines.  These redshifts are provided in Table~\ref{tab:samp}, and are mostly accurate to four decimal places.  We discuss these in detail in Section~\ref{sec:res}.

Based on radio morphology and spectral energy distribution (SED) at cm wavelengths, 
the associated targets can be classified as compact radio sources ($<$15kpc) expanding through the host galaxy ISM \citep[][]{Odea21}.
The expected values of radio continuum at 3mm are taken from the NASA Extragalactic Database (NED).   As many AGNs have variable fluxes, these are certainly only approximate. 
The flux densities from our mm observations, which are presented in the following tables, are also
uncertain because these are obtained with a single dish and through a varying atmosphere. 
The atmospheric contribution is best subtracted though rapid wobbling; however, it remains problematic 
in average weather conditions. 

We searched for molecular absorption at the lowest possible frequency to maximize the continuum brightness level, and hence the detectability of absorption.  All sources have been observed in CO lines and according to their redshift HCO$^+$ lines when possible. Our goal is to detect CO in emission or absorption and high-density tracers  simultaneously, such as HCO$^+$ or HCN, which become diffuse gas tracers in absorption.  
We note that some of the associated absorbers from the sample have been followed up to search for the OH line in absorption, namely with MeerKAT through the approved large program - the MeerKAT Absorption Line Survey \citep[MALS\footnote{$https://mals.iucaa.in/$};][]{Gupta2016, Gupta2021}.

\begin{table*}
\caption{Properties of the sample -- first 16 associated systems and then 14 intervening absorbers are presented.  
  Columns: (1,2) object coordinates and favored name;
  (3,4) redshift of emission and absorption, the latter is based on the \hi\ 21cm absorption peak, except for PKS1200+045 (see text for details);
  (5) observed frequency, CO(1-0) or CO(2-1) except when indicated otherwise;
(6) 3mm continuum flux density in mJy based on the measurements from NED;
        (7) epoch of IRAM 30m observations;
  (8) \hi\ 21cm absorption references. \\
}
\vspace{-0.4cm}
\begin{center}
\begin{tabular}{llcccccl}
\hline
\hline
Target   &  Other name &  $z_{em}$  & $z_{abs}$ & $\nu_{obs}$(1) & S$_{cont}$ & Epoch & \hi\ 21-cm references  \\
   &   &    &  & (GHz) & (mJy) &    &   \\
   (1) & (2) & (3) & (4) & (5) & (6) & (7) & (8) \\
   \hline
 J000557.17+382015.1&B0003+38A   &  0.229 &  0.2288 & 93.808   & 900 &05-19 &\citet{Aditya2018}\\
 J010826.84$-$003724.1& UM305    &  1.3753& 1.3710 & 97.232   & 200 &04-21 &\citet{Gupta2007}\\
                    &            &        &       &112.846$^a$&    &07-10 &  \\
 J014652.79$-$015721.2& 4C-02.08 & 0.9590 &  0.9589 & 91.059$^a$& 20 &05-19 &\citet{Aditya2019}\\ 
 J022928.93+004429.5&J0229+0044  & 1.2161 &1.2166 & 104.008  &0.5  &04-22 &\citet{Chowdhury2020}\\
 J022947.23+005308.9&J0229+0053  & -      & 1.1630 & 106.582  & 2   &04-22 &\citet{Chowdhury2020}\\
 J043103.76+203734.2&PKS0428+20  &0.219   & 0.2202  & 94.469   &300  &07-10 &\citet{Vermeulen2003}\\ 
 J104830.37+353800.8&B2 1045+35A &0.8464  &  0.8471 &124.811   & 20  &04-21 &\citet{Aditya2019}\\ 
 J120321.93+041419.1&PKS1200+045 & 1.2243 &   1.2111 &104.263   &200  &04-21 &\citet{Aditya2018}\\ 
 J124823.89$-$195918.8&PKS1245-19&1.275   &1.2750 &101.335   &20   &04-21 &\citet{Aditya2018}\\ 
 J132616.51+315409.5& 4C+32.44    &0.3680  &   0.3680 & 84.263   &1000 &07-10&\citet{Vermeulen2003}\\
 J134733.36+121724.2& 4C+12.50    &0.1217  &0.1217 &102.765   &1500 &07-10 &\citet{Morganti2004}\\
 J140700.39+282714.7&  Mrk668    & 0.0768 &0.0775 &106.980   &250  &07-10 &\citet{Gupta2006}\\ 
 J154015.23$-$145341.9&J1540-1453& 2.104  &    2.1139 & 74.035   & 60  &04-21 &\citet{Gupta2021}\\ 
 J164801.53+222433.3&J1648+2224  & 0.8227 &   0.8233 &126.440   &100  &04-21 &\citet{Aditya2019}\\ 
 J194553.51+705548.7&  1946+708  &0.1008  & 0.1008 &104.716   &48   &06-10  &\citet{Peck1999} \\ 
 J205252.05+363535.3&B2-2050+36  &0.354   &  0.3546 &85.096    &214  &06-10 &\citet{Vermeulen2003}\\ 
 \\
 \hline
 \\
 J020346.66+113445.4&PKS0201+113 &3.639   &  3.38714  &131.354  & 100 &07-10&\citet{Kanekar2007}\\ 
                    &.           &        &        &105.089     &     &07-10&     \\
 J025134.54+431515.8& Q0248+430  & 1.313  &  0.05151 & 109.624  & 150 &07-10&\citet{Hwang2004} (1)\\ 
 J074110.70+311200.2&0738+313    &0.6310  &   0.22124 &94.389   &800  &06-10&\citet{Kanekar2001b0738},(2)\\ 
 J080839.66+495036.5&SBS0804+499 & 1.4344 &   1.40732  &95.765  &200  &03-19,07-10&\citet{Gupta2009}, (3)\\ 
 J083052.09+241059.8&0827+243    & 0.9406 &   0.52476 &75.600   &900  &06-10&\citet{Kanekar2001}, (2)\\ 
 J092136.24+621552.2&J0921+6215  &1.4473  &  1.10360 &109.592   &1000 &03-19&\citet{Dutta2017}, (4)\\ 
 J095456.82+174331.2&0952+179    & 1.4749 &  0.23780 &93.126    &40   &06-10&\citet{Kanekar2001}\\
 J124355.79+404358.9&B3 1241+410 &1.5266  &  0.01714 & 113.329  & 80  &07-10&\citet{Gupta2018}\\ 
 J133335.78+164903.9&HB1331+170  &2.0835  &    1.77646 &83.033  &500  &03-19 &\citet{Carswell2011}, (2)\\ 
 J140856.48$-$075226.5&PKS1406-076&1.494  &   1.27464 &101.351  &1500 &03-19,06-10&\citet{Gupta2012}\\ 
 J162439.09+234512.2& 3C336      &0.9272  &   0.65584 &107.725$^a$&100 &06-10&\citet{Curran2007}\\ 
 J163956.36+112758.7& J1639+1127 & 0.993  &   0.07909 & 106.823  &70   &07-10 &\citet{Srianand13dib}\\ 
 J203155.23+121940.4&PKS2029+121 &1.215   &   1.11614 &84.293$^a$&2000 &06-10&\citet{Gupta2012}\\
 J235810.87$-$102008.7&HB2355-106&1.6349  &   1.17304 &106.090   &600  &07-10&\citet{Gupta2007}, (3)\\ 
\hline 
\end{tabular}
\tablefoot{ 
\tablefoottext{a}{for HCO$^+$(2-1) or (3-2)}    
 -- Other references (1) \citet{Combes2019}, (2) \citet{Berg2015},
 (3) \citet{Quider2011}, (4) \citet{Stickel1993} }
\label{tab:samp}
\end{center}
\end{table*}

\section{Observations and data analysis}
\label{sec:obs}

All the targets were first observed with the IRAM-30m telescope at Pico Veleta,
Granada, Spain, at the epochs indicated in Table~\ref{tab:samp}.
The lines observed were mainly either CO(1-0) or CO(2-1) at 3mm and  CO(3-2) at 2mm,
except in some cases HCO$^+$(2-1) or (3-2). The full width at half maximum (FWHM) of the primary beams
at the frequencies of 100\,GHz and 150\,GHz are 25$^{\prime\prime}$ and 17$^{\prime\prime}$, respectively. 
The SIS receivers (EMIR) were
used for observations in the wobbler switching mode, with reference
positions offset by $\pm$60\,arcsec in azimuth. The 
efficiency of the main beam of IRAM is $\eta_{mb}$ = T$_A^*$/T$_{mb}$ = 0.84 and 0.78 at 100 GHz
and 150 GHz, respectively. The system temperatures ranged 
between 120\,K and 220\,K at 3 mm, and between 180\,K and 300\,K 
at 2 mm. The pointing accuracy was checked every 2 h on
a nearby planet or a bright continuum source, and the focus was
reviewed after each sunrise, as well as at the beginning of each night.
The integration time is typically 2h per source, and is weather-dependent. 
Two backends were used simultaneously, the wideband autocorrelator, WILMA, 
and the Fourier Transform Spectrometer (FTS). Their respective spectral resolutions 
are 2 MHz and 0.2 MHz, which correspond to 6 and 0.6 \kms\ at 3mm. The rms noise
levels  at 100 GHz and 150 GHz for a velocity resolution of 40\,\kms\ 
were $\sigma \sim$1.0\,mK and 1.3\,mK in T$_A^*$, respectively. 
The data were reduced using the Continuum and Line Analysis Single-dish Software (CLASS) of 
the Grenoble Image and Line Data Analysis Software (GILDAS).

For two sources detected in line emission, \pks\ and \pkss, further observations in the 3mm band 
were carried out with NOEMA using the PolyFIX correlator in 
February, March, and November, 2022. The CO(2-1) line at $z$ = 1.226 (\pks)
and 1.275 (\pkss) is redshifted to 103.566 and 101.335 GHz, respectively. 
The observations were carried out in dual polarisation mode in four base bands, 
with 3.9 GHz total bandwidth per baseband, distributed
in lower and upper sidebands distant by 15.5~GHz. The CO(2-1) line was observed in the upper side band.
We observed 80\% of the time in A-configuration, 
and 20\% in C-configuration, with 12 antennas. The total telescope time per source 
was 13h (8h on source), with a total of 26h. 
The calibrations were performed using strong radio sources, such as 3C273, 3C84, 1222+037, and
1244-255. The absolute flux calibration is accurate at the 10\%
level. 

The data were calibrated using the Continuum and Line Interferometer Calibration (CLIC) 
package and mapped using the MAPPING package of the GILDAS software.
We used self-calibration, which significantly increased  the signal-to-noise
ratio. Using CLARK cleaning with natural weighting, we obtained images with 
synthesized beams of 6$\farcs$9$\times$1$\farcs$3 (\pks) and 4$\farcs$1$\times$1$\farcs$2 (\pkss)
with a position angle (PA) of 10\textdegree.
As no other lines were detected in the rest of the base bands (upper and lower), we 
used these to estimate and subtract the continuum level. 
The velocity resolution was initially 2\,MHz ($\sim$6\,\kms). The data were then
smoothed to obtain data cubes with three resolutions: 10, 50, and 150\,\kms. 
These final cubes are 256$\times$256 pixels with 0.13\arcsec per pixel in the plane
of the sky, and have 80 frequency channels. 
The continuum flux density was also computed using the images from the wider 7.7\,GHz lower
side band, with a beam of 7$\farcs$1\arcsec$\times$1$\farcs$4 or
4$\farcs$9$\times$1$\farcs$4, with a PA of 10\textdegree.
The rms noise is 230\,$\mu$Jy/beam in 50 \kms\ channels for the line and 10\,$\mu$Jy/beam for the continuum.

\begin{figure}
\begin{center}
\includegraphics[clip,width=0.51\textwidth,angle=0]{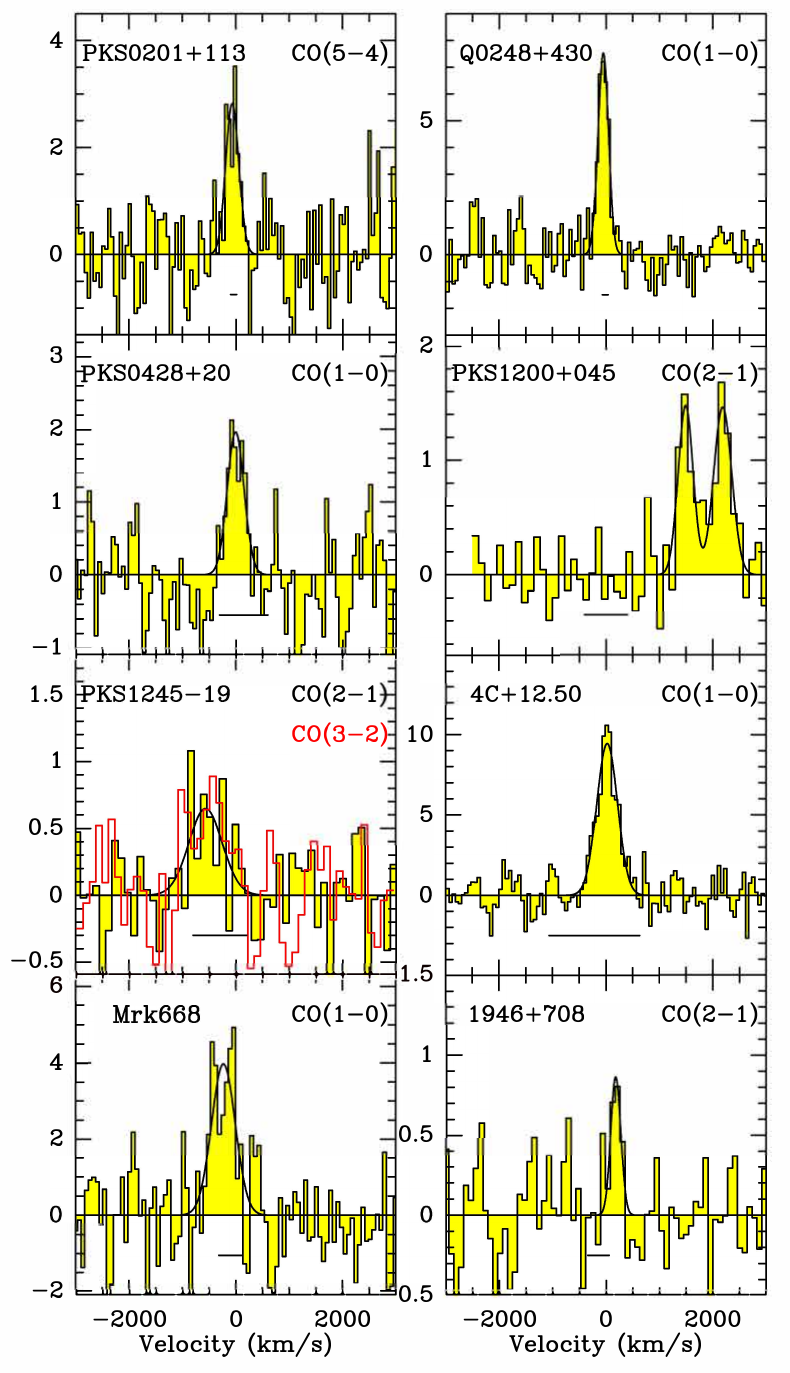}
\vskip+0.0cm
\caption{ Detection of CO emission lines toward two intervening (top panels) 
and six associated absorbers obtained with the IRAM-30m.
The vertical scale is T$_{mb}$ in mK. The various 
CO transitions are indicated at the upper right of each panel.  The zero of the velocity scale corresponds to $z_{abs}$ in Table~\ref{tab:samp}. The velocity range over which \hi\ 21cm absorption is detected is indicated by an horizontal bar at the bottom of each panel.
The detection toward 1946+708 is only tentative.} 
\label{fig:em8} 
\end{center}
\end{figure} 

\section{Results}
\label{sec:res}

In Section~\ref{sec:resiram30m}, we first discuss the emission and then absorption line detections 
from IRAM-30m observations. The results from NOEMA interferometer follow-up observations of associated systems \pks\ and \pkss\ are presented in Section~\ref{sec:resnoema}.

\subsection{IRAM-30m}
\label{sec:resiram30m}

\begin{figure}
\begin{center}
\includegraphics[clip,width=0.49\textwidth,angle=0]{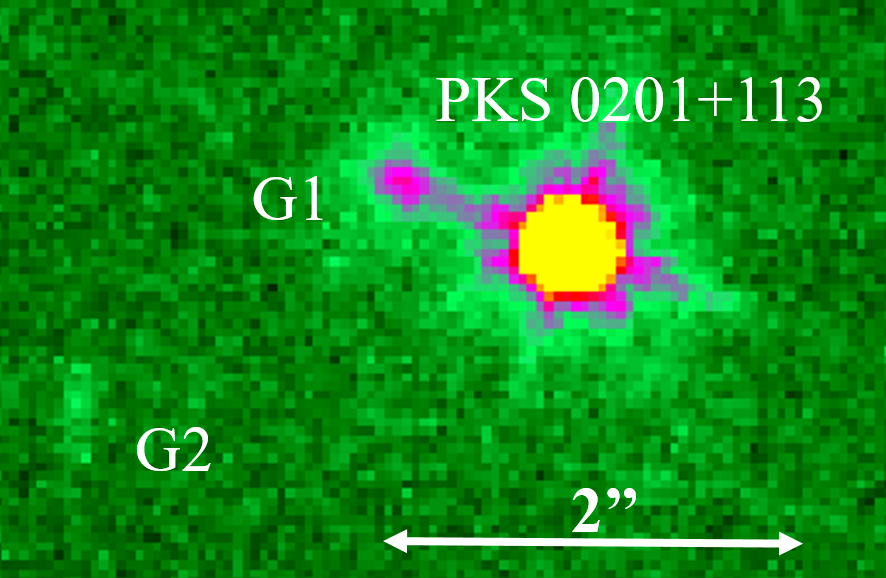}
\includegraphics[clip,width=0.49\textwidth,angle=0]{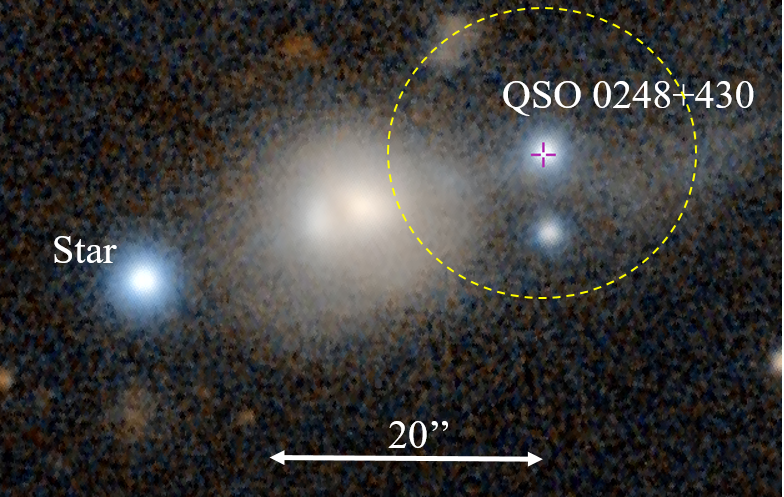}
\vskip+0.0cm
\caption{QSO-galaxy pairs detected in CO emission and absorption. Top: HST-F606W image of PKS0201+113, with the galaxies G1 and G2; the latter is considered the most likely DLA candidate by \citet{Ellison2001}. 
Bottom: PanSTARRS1  three-color (z-zg-g) image of Q0248+430
and associated galaxies. The dashed yellow circle indicates the region observed with the 23\arcsec beam.} 
\label{fig:QSO-G} 
\end{center}
\end{figure} 

\subsubsection{Emission}

Figure~\ref{fig:em8} displays profiles of CO emission lines detected toward eight of  
the targets with the IRAM-30m telescope. 
 We note that the detection toward 1946+708 is only tentative.
Of these eight, two correspond to intervening \hi\ 21cm absorbers 
and the remaining to the associated absorbers (see also Table~\ref{tab:emassoc}).
The detections toward Q0248+430 \citep[J0251+4315;][]{Downes1993}, 4C+12.50  
\citep[J1347+1217;][]{Dasyra2012} and Mrk668 \citep[J1407+2827;][]{Ocana2010} were previously known, and the remaining are reported here for the first time.  

The two intervening systems detected in CO emission are toward PKS0201+113 (J0203+1134) and  Q0248+430 (J0251+4315).  Although both can be classified as QSO-galaxy pairs (QGP;
i.e., a foreground galaxy producing absorption in the sight line of a distant quasar as shown in Fig.~\ref{fig:QSO-G}), the absorber toward PKS0201+113 was first identified as a DLA at $z_{abs}$ = 3.38639 \citep[][]{White1993} and was then subsequently associated to a foreground concentration of galaxies, with one of them producing the absorption \citep[][]{Ellison2001}. The absorber galaxy responsible for the DLA may be G2 in Fig.~\ref{fig:QSO-G}.  It is relatively massive (0.7$L^*$) and at an impact parameter of $\sim$20\,kpc. The DLA with N(\hi) = $10^{21.26\pm0.08}$\,\cmsq\ exhibits multiple components spread over $\sim$270\,\kms.   The \hi\ 21cm absorption exhibits two components spread over $\sim$115\,\kms\ at $z_{abs}$ = 3.387 144(17) and 3.386 141(45) \citep[][]{Kanekar2007}. The $z_{abs}$ adopted in Table~\ref{tab:samp} corresponds to the peak of stronger 21cm absorption component.  The 21cm absorption peaks do not coincide with the metal and H$_2$ absorption lines detected in the optical/UV spectra \citep[][]{Ellison2001, Kanekar2007, Srianand2012}. The CO emission spectrum is in good correspondence with the velocity range over which [FeII], [CII] and H$_2$ absorption lines are detected \citep{Srianand2012}.

Being closer ($z \sim$  0.05), the merging galaxy pair in front of the quasar Q0248+430 (J0251+4315) at $z$ = 1.313 has been detected in several CO emission lines \citep[][]{Kuehr1977, Kollatschny1991}. The CO(1-0) integrated flux density ($S_{CO}dV$) using the Berkeley-Illinois-Maryland Association (BIMA) interferometer was 24\,Jy\,\kms\ \citep{Hwang2004}. With the IRAM-30m, \citet{Downes1993} reported $S_{CO}dV$ = 25\,Jy\,\kms, while from our observations we estimate $S_{CO}dV$= 8.5\,Jy\,\kms.  However, in our observations, the telescope was pointed toward the QSO, which is about 15\,\arcsec\ from the foreground galaxy.  \citet{Combes2019}, using NOEMA observations, showed that the CO emission is confined to the few central arcseconds of the galaxy. Therefore, the telescope beam only partially covers the CO emission region associated with the galaxy, as shown in Fig. \ref{fig:QSO-G}, which explains the discrepancy with earlier observations. We note that the tidal tail emanating from the foreground galaxy extends across the background QSO and is responsible for the CO absorption line detection \citep[][]{Combes2019}.  Unlike the case of PKS0201+113, the molecular absorption in this case is coincident with the narrower and stronger \hi\ 21cm absorption components \citep[see also][]{Gupta2018oh}. 

\begin{figure}
\begin{center}
\includegraphics[clip,width=0.49\textwidth,angle=0]{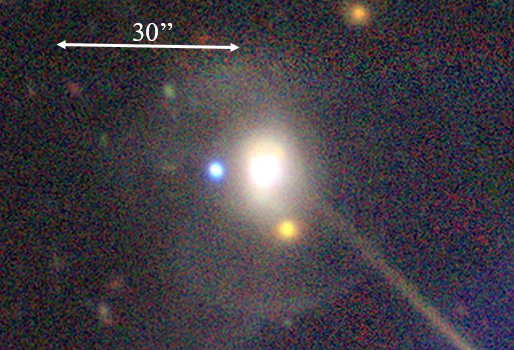}
\vskip+0.0cm
\caption{Mrk668 (OQ208, J1407+282) legacy DR9 image, showing an interacting 
galaxy with stellar shells.} 
\label{fig:OQ208} 
\end{center}
\end{figure}

Among six associated molecular-emission-line detections, the four lower-redshift detections are associated either with radio or Seyfert (Sy) galaxies (see Table~\ref{tab:emassoc}). 4C+12.50 (J1347+1217) is associated with a merging galaxy pair at $z$ = 0.12174 $\pm$ 0.00002 \citep[][]{Holt2003}, with infrared luminosity of 3.2$\times10^{12}$\,L$_\odot$ \citep[][]{Scoville2000}. This has been detected in CO in emission as well as absorption \citep[][]{Dasyra2012,Dasyra2014}. An image of the associated galaxy pair can be seen in Fig.~5 of \citet{Dasyra2014}. 
The host of Mrk668 (J1407+2827; OQ208) is an actively star forming galaxy, with an interacting companion (see Fig. \ref{fig:OQ208}),  at $z$ = 0.07681 $\pm$ 0.00007 \citep[][]{Eracleous2004}. It was detected previously by \citet{Ocana2010} in CO(1-0) and also shows stimulated radio-recombination lines \citep{Bell1980}. Due to its far-infrared luminosity of 2$\times$10$^{11}$ L$_\odot$, it is classified as a LIRG. 
1946+708 (J1945+7055) is a radio galaxy at $z$ = 0.10083 $\pm$ 0.00009 \citep[][]{Snellen2003}.  
PKS0428+20 (J0431+2037), also a radio galaxy, has a redshift measurement of $z$ = 0.219 with modest accuracy \citep[][]{Labiano2007}.

The two higher-redshift associated detections are PKS1200+045 (J1203+0414) and PKS1245-19 (J1248-1959) at $z$ = 1.22429 $\pm$ 0.00043 \citep{Shen2011,Aditya2018} and 1.275 \citep[][]{Labiano2007}, respectively.  For PKS1200+045, the double-peaked CO emission profile agrees with the optical redshift within 1$\sigma$.  The \hi\ 21cm absorption is blueshifted with respect to this by $\sim$2000\,\kms\ and has no overlap with the molecular gas.  In contrast, for PKS1245-19, the \hi\ 21cm absorption and CO emission line peaks are consistent within the random and systematic errors of measurements with the systemic redshift based on optical emission lines.  In general, except for PKS1200+045, for all the associated systems, the \hi\ 21cm absorption is detected within the same velocity range as the CO emission (see horizontal bars in Fig.~\ref{fig:em8}).

\begin{table*}
\caption{Detection of CO emission in two intervening systems, PKS0201+113 and Q0248+430, and six associated systems. Columns: (1): favored name; (2) $z$ of the CO emission peak; (3) CO transition; (4, 5): emission line peak and FWHM; (6): molecular mass; (7, 8): optical and radio identifications  (GPS: GHz-peaked spectrum, CSS: compact steep spectrum), and 1.4\,GHz luminosity for associated systems.}
\vspace{-0.4cm}
\begin{center}
\begin{tabular}{lcccccccc}
\hline
\hline
 Target    & $z_{mol}$  & CO  &S$_{peak}$ &  FWHM  & M(H$_2$)  & Optical Id. & Radio Id.  & log$L_{1.4\,GHz}$\\
           &    &     &   (mJy)     &  (\kms)   & (10$^9$M$_\odot$) & & & (W\,Hz$^{-1}$) \\
(1)        & (2) & (3) & (4) & (5) & (6) & (7) & (8) & (9) \\ 
 \hline   
 PKS0201+113&3.38715&5-4 &11$\pm$3   &289$\pm$50 & 768  &  -    &    -    &   -      \\
 Q0248+430  &0.05151&1-0 &38$\pm$5   &225$\pm$23 &  5   &  -    &    -    &   -      \\
 PKS0428+20 &0.22020&1-0 &10$\pm$2   &368$\pm$64 & 40   &  G    &    GPS  &  26.6    \\
 PKS1200+045&1.22216&2-1 &6.4$\pm$1  &750$\pm$90 & 488  &  Q    &    GPS  &  27.9    \\
 PKS1245$-$19&1.27121&2-1&3.7$\pm$1  &752$\pm$140& 327  &  Q    &    GPS  &  28.6    \\
             &1.27121&3-2&3$\pm$1.5  &733$\pm$180& 190  & \\
 4C+12.50    &0.12170&1-0 &47$\pm$5   &457$\pm$32 & 71   & Sy/Q  &    CSS  & 26.3     \\
 Mrk668     &0.07642&1-0 &20$\pm$5   &521$\pm$69 & 13   & Sy    &    GPS  &  24.9    \\
 1946+708   &0.10147&2-1 &4.3$\pm$1  &222$\pm$90 & 0.6  & G     &    GPS  &  25.3    \\
\hline
\end{tabular}
\label{tab:emassoc}
\end{center}
\end{table*}

We compute L$^\prime_{CO}$, the CO luminosity in  units of K \kms pc$^2$,
with the integrated emission in the beam. This CO luminosity is given by
\begin{equation}
L'_{CO} =
3.25 10^7 S_{CO}dV  {{D_L^2}\over {\nu_{rest}^2(1+z)}} \hskip6pt \rm{K\hskip3pt
  km \hskip3pt s^{-1}\hskip3pt pc^2},
\end{equation}
where  $S_{CO}dV$ is the integrated flux  in Jy \kms\,, $\nu_{rest}$ is the rest
frequency in GHz,  and $D_L$ is the luminosity distance in megaparsecs.
Under the  assumption of a standard CO-to-H$_2$ conversion factor \citep{Bolatto2013},
we compute the H$_2$ mass using M$_{\rm H_2} = \alpha$ L'$_{\rm CO}$,
with $\alpha=4.36$ M$_\odot$ (K \kms\, pc$^2$)$^{-1}$.  We use a constant
conversion factor for the sake of comparison, although the factor could depend on the
starbursting character and on the metallicity of galaxies, but we have no information 
about them.
The observation of the fundamental CO(1-0)
line is the best measure of the total \hh\, mass.
To convert the CO luminosities measured at a $J$-level of higher than one, we adopted the 
$R_{1J} = T_1/T_J$ ratios from several works in the literature
\citep[e.g.][]{Weiss2007,Dannerbauer2009,Tacconi2018}, i.e. $R_{12} = 1.16$, $R_{13} = 1.8$,
$R_{15} = 2.9$.  The molecular gas mass estimates for eight detections are presented in Table~\ref{tab:emassoc}.

\begin{figure}
\begin{center}
\includegraphics[clip,width=0.49\textwidth,angle=0]{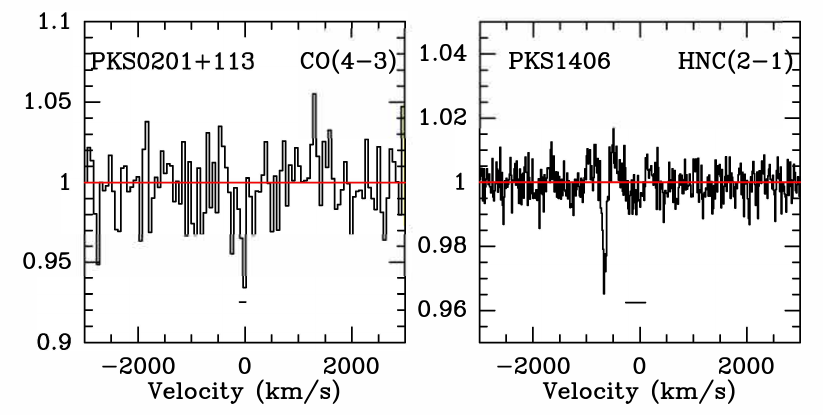}
\vskip+0.0cm
\caption{ Absorption features observed with the IRAM-30m toward two of the intervening targets -- PKS0201+113 in CO(4-3) on the left, and PKS1406-076 in HNC(2-1) on the right. 
The detection toward PKS0201+113 is only tentative.
The spectra have been normalized to their continuum flux densities of 0.22 and 1.13\,Jy, respectively. The zero on the velocity scale corresponds to the peak of \hi\ 21cm absorption ($z_{abs}$ in Table~\ref{tab:samp}). The velocity range over which \hi\ absorption is detected is indicated by a horizontal bar at the bottom of each panel. For PKS0201+113, the H$_2$ absorption peak (-25\,\kms) and the velocity range over which the 21cm (-100, 10)\,\kms\ and the [CII]$^*$ (-80, 0)\,\kms\ are detected are reasonably consistent with the CO absorption (see text for details).} 
\label{fig:abs2} 
\end{center}
\end{figure} 

\subsubsection{Absorption}

Figure \ref{fig:abs2} shows the spectra of the two 
molecular-absorption-line detections. They are both intervening systems.
The values of continuum flux density, optical depth, and column density are provided in Table~\ref{tab:absinter}. 
PKS0201+113 exhibits a weak absorption ($\sim$3$\sigma$) in the CO(4-3) line, 
 and the detection is only tentative.
The absorbing galaxy is also detected in emission in CO(5-4) (see Fig.~\ref{fig:em8}). The continuum flux density of the quasar is a factor of two higher at the lower frequency of the CO(4-3) line with respect to the higher frequency of CO(5-4). At the latter frequency, the absorption signal is weak and does not compensate for the emission. The contrary is true at the CO(4-3) frequency: the absorption signal is stronger and cancels out the emission.   The CO(4-3) absorption coincides well with the velocity range (-100, 10)\,\kms\ over which the \hi\ 21cm absorption is detected.  However, we note that the peaks of 21cm and H$_2$ absorption lines, despite both being tracers of cold gas, are disparate by $\sim$25\,\kms\ \citep[][]{Srianand2012}.  The differences in absorption line velocities are most likely due to the nonalignment of radio and optical sight lines.  The radio core, especially at cm wavelengths, is also more extended than the astronomical unit(au)-sized optical quasar and probes a larger volume of the absorbing gas.

PKS1406-076 was observed in June 2010 at 101.322\,GHz and 151.978\,GHz corresponding to the CO(2-1) and CO(3-2) lines redshifted at $z$=1.2753, respectively.  The \hi\ 21cm absorption peak is at $-$87\,\kms\ with respect to $z$=1.2753, corresponding to the Mg~{\sc ii} absorption.  The observation led only to upper limits ---with a spectral rms of 8\,mJy per channel of 20\kms--- along with rippled baselines, which are likely due to the high continuum flux density (1.1\,Jy). In March 2019, the target was observed again but only in the 3mm band with a wide configuration, which included CO(2-1) at 101.322 GHz in the upper-outer sideband, but also HNC(2-1) and HCO$^+$(2-1) at 79.693\,GHz and 78.396\,GHz in the lower-outer sideband, respectively. An absorption line was detected near the HNC(2-1) frequency but blueshifted by $\sim$600\,\kms\ with respect to the peak of the 21cm absorption. 
The HCO$^+$(2-1) frequency was at the edge of the band and suffered from a rippled baseline.

The continuum flux densities and upper limits for nondetections are presented in Table~\ref{tab:upper}.  
The spectral rms ($\sigma$) is estimated for 20\,\kms\ channels. When the continuum is insufficiently strong, 
that is, regarding the CO line in particular, the emission line upper limit may be estimated using the rms 
derived over 80\,\kms channels obtained by dividing $\sigma$ in Table~\ref{tab:upper} by a factor 2. 
We do not report all the simultaneously observed lines, but only those offering the greatest constraints are considered here.

Assuming a homogeneous and optically thin gas at excitation temperature 
$T_{x}$ and with a covering factor $f_c$, we can derive the average column density
over the beam corresponding to the mm continuum emission of the quasar through
\begin{equation}
 N_{tot} = \frac{8\pi}{c^3} \frac{\nu^3}{g_J A_{J,J+1}} f(T_x) \int{\tau dv},
\end{equation}

\noindent where $g_J$ is the statistical weight of level $J$, $A_{J,J+1}$ is the Einstein coefficient for
transition $J \rightarrow J + 1$, and the function $f (T_x )$ is
\begin{equation}
f (T_x ) = \frac{Q(T_x ) e^{E_J /kT_x}}{1 - e^{-h\nu/kT_x}}
.\end{equation}
We  adopt the partition function of local thermal equilibrium (LTE),
that is,  $Q(T_x ) = \Sigma g_J e^{-E_J /kT_x}$,
where $E_J$ is the energy of level $J$ and $T_x$ is the
excitation temperature of the CO molecule.

Typically, with our integration time of 2h per source, 
  at 3mm we get an rms of 5\,mJy in 20 \kms channels. According to the 
background radio source flux, between 100 and 300\,mJy of continuum,
this leads to a 1$\sigma$ upper limit of $\sim$ 1.7$\times$ 10$^{-2}$  to 5$\times$ 10$^{-2}$ 
in the optical depth $\tau$, and $\tau \Delta$V = 0.34 to 1 \kms\ at 1$\sigma$
for a line width of $\Delta$V = 20 \kms.
When the observed line is CO(2-1) or HCO$^+$(2-1), this corresponds to
N(CO) = 1.6-5$\times$10$^{15}$ cm$^{-2}$, and N(H$_2$) = 1.6-5$\times$10$^{19}$ cm$^{-2}$,
or N(HCO$^+$) = 2.6-8$\times$10$^{12}$ cm$^{-2}$ respectively, 
assuming an average excitation temperature of $T_{x}$= 15K, and a filling factor of  $f_c$ = 1.  We adopt a typical CO/H$_2$ abundance ratio of $10^{-4}$ \citep[e.g.,][]{Rachford2009}.

\begin{table}
\caption{Details of molecular-absorption-line detections in two intervening and then two associated systems with IRAM-30m and NOEMA, respectively. Columns: (1) favored PKS name of the target; (2)  $z$ of the molecular line peak; (3) molecular line transition; (4) continuum flux density; (5) integrated optical depth; and 
(6) total column density of CO or HNC.}
\vspace{-0.4cm}
\begin{center}
\begin{tabular}{lccccc}
\hline
\hline
Target & $z_{mol}$& Line &S$_{cont}$  & $\int\tau$dV & $N$ \\
       &          &      &   (Jy)     &   (\kms)     & (\cmsq)\\
 (1)   & (2)      & (3)  & (4)        & (5)          & (6) \\
 \hline          
  0201+113   &3.38715 &CO43 & 0.22 & 6.2$\pm$2  & 6.6E16\\
  1406$-$076 &1.26949 &HNC21& 1.13 & 2.4$\pm$0.2& 2.8E13\\
  1200+045   &1.21276 &CO21 & 0.058& 3.1$\pm$0.6& 1.5E16\\
  1245$-$19  &1.26605 &CO21 & 0.067& 7.2$\pm$0.4& 3.6E16\\
\hline
\end{tabular}
\label{tab:absinter}
\end{center}
\end{table}

The column-density estimates for the two intervening detections with IRAM-30m 
are provided in Table~\ref{tab:absinter}. We note that, although the column density does not depend
on redshift, the excitation temperature must nevertheless be higher than the CMB temperature,
which is already 12\,K for PKS0201+113. When the assumed T$_x$ is 20K, instead of 15K,
the column density decreases only slightly (13\%) for CO(4-3), and increases by 53\% for HNC(2-1).
For a typical CO/H$_2$ abundance ratio of 10$^{-4}$, the 
H$_2$ column density is $N$(H$_2$) = 6.6$\times$10$^{20}$ cm$^{-2}$ for PKS\,0201+113.
This is quite comparable to the \hi\ column density of 1.8$\times$10$^{21}$ cm$^{-2}$
derived by \citet{Ellison2001} and \citet{Kanekar2007}, from the Lyman-$\alpha$ and \hi\ 21cm, respectively.
For PKS1406-076, we can estimate the 
abundance of HCN and HNC in the diffuse medium of the Milky Way in absorption in front of 
background radio sources \citep{Liszt2001} as HCN/H$_2$ = 1.4$\times$10$^{-9}$ and HNC/HCN = 0.2. 
The H$_2$ column density is then N(H$_2$) = 10$^{23}$ cm$^{-2}$.

\subsection{NOEMA}
\label{sec:resnoema}

\begin{figure}
\begin{center}
\includegraphics[width=0.24\textwidth,angle=0]{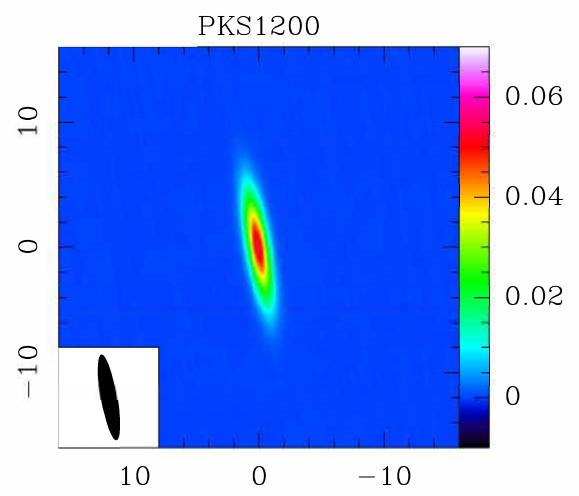}
\includegraphics[width=0.24\textwidth,angle=0]{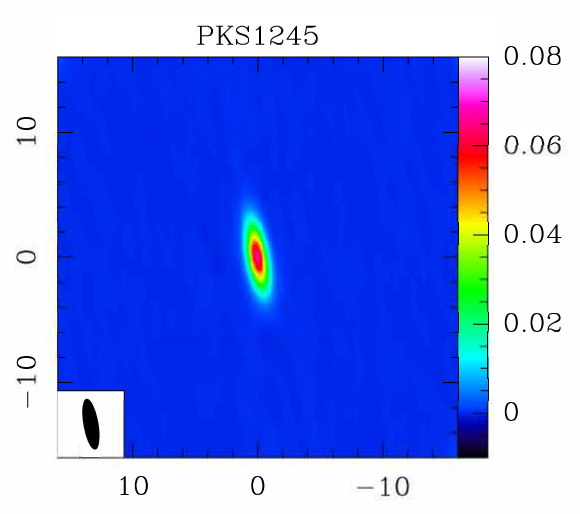}
\vskip+0.0cm
\caption{{\it Left:} Continuum image of PKS1200+045 at 3mm obtained with NOEMA. Only a 1.5 GHz band has been used around the line frequency to better subtract the continuum. The synthesized beam is 6$\farcs$9$\times$1$\farcs$3, with PA=10\textdegree. The compact source has a flux density of 58\,mJy and detected at S/N of 303.
{\it Right:} Same as above  but for PKS1245-19. The synthesized beam is 4$\farcs$1$\times$1$\farcs$2, with PA=10\textdegree. 
The compact source has a flux density of 67\,mJy (S/N = 421).
In both the panels, the spatial scale is in arcseconds and the color bar is in Jy\,beam$^{-1}$.
} 
\label{fig:PKScont} 
\end{center}
\end{figure}

\begin{figure}
\begin{center}
\includegraphics[clip,width=0.5\textwidth,angle=0]{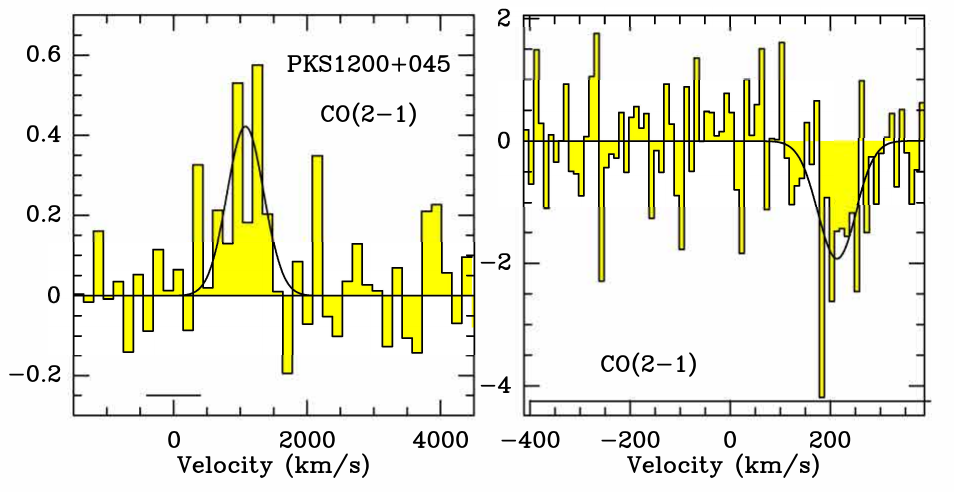}
\vskip+0.0cm
\caption{{\it Left:} CO(2-1) emission spectrum of PKS1200+045 obtained with NOEMA and 
smoothed to 150\kms. The velocity scale is centered at the average of  \hi\ absorption.
{\it Right:} CO(2-1) absorption spectrum toward PKS1200+045. The spectral resolution is 10\,\kms.
In both panels, vertical scales are in mJy\,beam$^{-1}$ and the horizontal line at the bottom indicates the velocity range over which the \hi\ 21cm absorption is detected \citep[][]{Aditya2018}.} 
\label{fig:PKS1200} 
\end{center}
\end{figure}

\begin{figure}
\begin{center}
\includegraphics[clip,width=0.35\textwidth,angle=0]{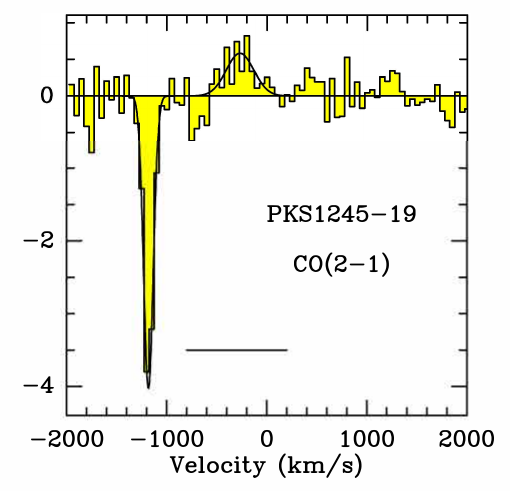}
\vskip+0.0cm
\caption{CO(2-1) emission and absorption spectrum of PKS1245-19 obtained with NOEMA and 
smoothed to 150\,\kms. The vertical scale is mJy\,beam$^{-1}$. 
The horizontal line at the bottom indicates the velocity range over which the \hi\ 21cm absorption is detected \citep[][]{Aditya2018}.
} 
\label{fig:PKS1245} 
\end{center}
\end{figure}

The two sources \pks\ and \pkss\ are clearly detected in 3mm continuum emission (see Fig. \ref{fig:PKScont}). 
The CO emission lines are also detected, as shown in Figs.~\ref{fig:PKS1200} and  \ref{fig:PKS1245}, 
but with much lower amplitude than in the IRAM-30m spectra.
Also for \pks\, the line shape is different and is blueshifted by
$\sim$ 1000 \kms; it is therefore doubtful.
The difference might also be due to extended CO emission being resolved out by
the limited uv-coverage of the interferometer. However, as the smallest extent of the synthesized beam is $\sim$ 1\arcsec or $\sim$ 8~kpc, this seems unrealistic. Another possibility is that the dynamic range
 is limited to detect weak line emission in the presence of a strong continuum source.
Indeed, the intensity of the latter was found to vary significantly within the baseband, 
affecting the accuracy of the subtraction of the continuum.
Computed over the lower and upper sidebands ($\sim$ 20 GHz),
the spectral index of the source is close to -1; -0.6 for \pks\ and -1.5 for \pkss.

We also detect the CO(2-1) absorption lines in both sources. At V=-1803\,\kms 
with respect to the optical redshift, the 
molecular absorption is consistent with the velocity range over which 
the \hi\ 21cm absorption 
is detected by \citet[][]{Aditya2018}.  The absorption peaks are shifted by $\sim$200\,\kms\ and the molecular line overlaps with the weaker, shallower \hi\ absorption component. 
For \pkss, the molecular absorption line peak is at V= -1180\,\kms\ with respect
to the optical redshift, that is, the edge of the blueshifted wing of 
the \hi\ 21cm absorption \citep[][]{Aditya2018}.
Although the CO(2-1) absorption associated with \pkss\ appears very deep in the NOEMA spectrum, 
it is not incompatible with the nondetection in the IRAM-30m spectrum: the peak of the absorption is -4\,mJy,
and the spectral rms in the IRAM-30m spectrum is 2.5 mJy per 50\kms\ channel.

\begin{table}
\caption{Continuum flux densities and line upper limits based on IRAM-30m for the associated (top) and then intervening systems (bottom). The spectral rms ($\sigma$) is based on 20~\kms channels.}
\vspace{-0.4cm}
\begin{center}
\begin{tabular}{lcccc}
\hline
\hline
Target  & $\nu_{obs}$ & S$_{cont}$ & Line   &  $\sigma$ \\
        &  (GHz)       & (mJy)         &       &  (mJy) \\
   \hline
 B0003+38A   & 93.808  & 90  & CO(1-0)  & 7\\
 UM305       & 97.232  & 120 & CO(2-1)  &12 \\ 
             & 112.846 & 100 &HCO$^+$(3-2)& 15\\
 4C-02.08    & 91.059  & 20  &HCO$^+$(2-1)& 4.5\\
 J0229+0044  & 104.008 & 10  &CO(2-1)   &2.5\\
 J0229+0053  & 106.582 &  5  &CO(2-1)   &3.5 \\
 B2 1045+35A & 96.570  & 45  &HCO$^+$(2-1)&3.5\\
             & 124.811 & 30  &CO(2-1)   &6 \\ 
 4C+32.44     & 84.263  & 240 &CO(1-0)   & 9 \\
 J1540-1453  & 74.035  & 25  &CO(2-1)   & 7\\
 J1648+2224  & 126.440 & 175 &CO(2-1)   & 5\\
 B2-2050+36  & 85.096  & 80  &CO(1-0)   &2.5\\
 \\
 \hline
 \\
 0738+313    & 94.389  & 370 &CO(1-0)   & 3.5\\
 SBS0804+499 & 95.765  &440  &CO(2-1)   & 12\\
 0827+243    & 151.196 & 610 &CO(2-1)   & 5\\
 J0921+6215  & 109.592 & 850 &CO(2-1)   & 7\\
 0952+179    & 93.126  &200  &CO(1-0)$^a$& 3.5\\
 B3 1241+410 & 113.329 & 40  &CO(1-0)   & 10\\
 HB1331+170  & 83.033  &100  &CO(2-1)   &2.5\\
 3C336       & 107.725 &25   &HCO$^+$(2-1)& 5\\
             & 139.227 &15   &CO(2-1)   & 3.5\\
 J1639+1127   & 106.823 & 340 &CO(1-0)   & 10\\
 PKS2029+121 & 84.293  & 900 &HCO$^+$(2-1)& 4\\
             & 163.409 & 505 &CO(3-2)   & 5\\
 HB2355-106  & 106.090 &490  &CO(2-1)   & 4.5 \\
\hline 

\end{tabular}
\tablefoot{ 
\tablefoottext{a}{CO(4-3) detected by ALMA at z$_{em}$ \citep{Audibert2022}}    
 }
\label{tab:upper}
\end{center}
\end{table}


\section{Discussion}
\label{sec:disc}
 The rarity of molecular absorptions among systems with \hi\ 21cm absorption  (5 out of 30) ---even when the mm continuum is suffciently strong and the CO emission line is detected--- might appear surprising. But this is likely the consequence of the low surface filling factor (f$_s$) of the molecular gas. 
 There may be molecular gas along the line of sight, with a rather wide beam of 10\,kpc scales, as seen in the cases with CO emission, but the surface may be filled at f$_s\sim$1\% or lower \citep{Wiklind1997}. In contrast, the filling factor of the atomic gas is higher for the same column density. In addition, radio continuum emission is more extended at cm wavelengths than at mm wavelengths.   Both these factors lead to more favorable conditions for the detection of cold atomic gas in absorption at longer wavelengths than molecular gas at mm wavelengths.

 When the CO lines are detected in emission, higher-spatial-resolution observations are required to isolate the 
 compact nuclear continuum emission and avoid a situation where the emission line outshines the absorption signal. 
 NOEMA was used for two of the associated systems detected in emission. Although the continuum is strongly detected, the absorption signal is rather weak. The detection of weak absorption signals is made even more difficult by the influence of strong continuum emission on the noise level and the baseline stability. 
 Alternatively, it is possible to detect absorption in the high-density tracers, such as HCO$^+$, HCN, and HNC, which are not detectable in emission. This is indeed the case for the intervening system toward PKS1406-076. 

\begin{figure}
\begin{center}
\includegraphics[clip,width=0.45\textwidth,angle=0]{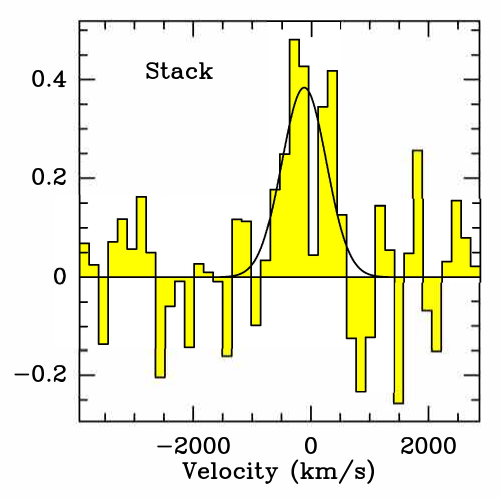}
\vskip+0.0cm
\caption{Spectrum stacked from of all upper limits described in Table \ref{tab:upper}.
The spectra were all aligned at the observed \hi\ absorption velocity. The vertical 
scale is T$_{mb}$ in mK. All continuum levels have been ignored.
} 
\label{fig:stack} 
\end{center}
\end{figure}

 The eight CO emission line detections in our sample correspond to $M$(H$_2$) = $10^9 - 7\times 10^{11}$\,M$_\odot$. Six of these are associated systems and the majority (i.e., four) are at low redshift ($z<$0.25).  Among intervening systems, one detection is at $z$=3.3871 and the other at 0.05151.
It is interesting to stack all the nondetection  spectra listed in Table~\ref{tab:upper} by aligning these to the same velocity scale centered at the peak of the \hi\ 21cm absorption. We stacked these spectra with no particular weight related to continuum level or redshift; simply taking into account their rms. The latter  was comparable for all sources, and therefore no individual one dominates.  The stacked spectrum in Fig. \ref{fig:stack} reveals a weak emission. The FWHM of the feature is rather large (870 \kms), and is most likely due to nonalignment of the \hi\ absorption with the systemic velocity of the galaxies. The integrated flux is 1.75$\pm$0.3 Jy\kms. We also stacked the associated and intervening nondetections separately, but the resulting spectra do not show any detectable feature.

\subsection{Comparison of \hi\ and H$_2$ column densities}

\begin{figure}
\begin{center}
\includegraphics[clip,width=0.45\textwidth,angle=0]{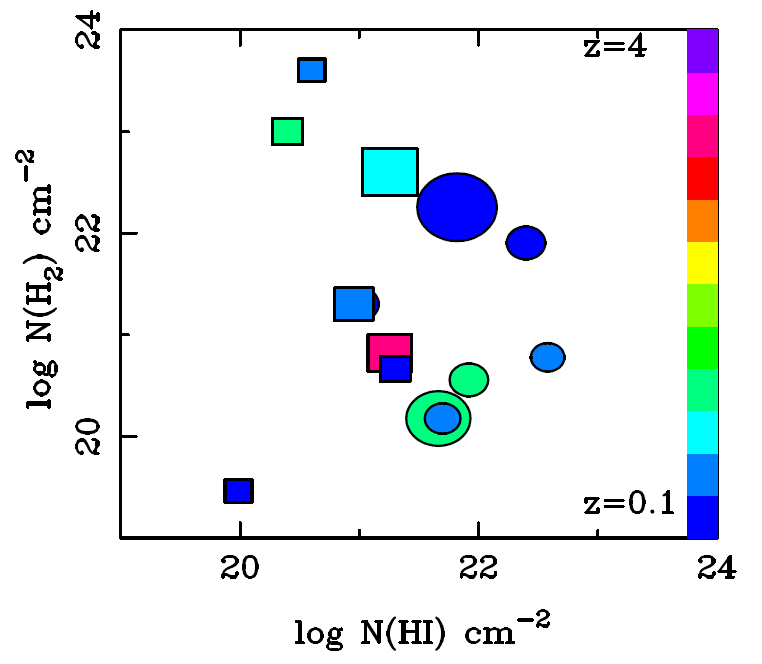}
\vskip+0.0cm
\caption{Comparison between H$_2$ and \hi\ (21cm) column densities 
for the known molecular-absorption-line systems at high redshift.  T$_s$ = 1000\,K (100\,K) is adopted for associated  (intervening) systems shown as circles (squares). The color of the symbols represents the redshift,
as indicated by the color bar on the right, and the size of the symbol traces
the velocity extent of the absorption averaged over the \hi\ and H$_2$ components. 
} 
\label{fig:compare} 
\end{center}
\end{figure}

We compared the atomic and molecular column densities of 14 known high-redshift ($z>0.1$) molecular-absorption-line systems. Table~\ref{tab:compare} provides a compilation of their basic properties from the literature.  Seven of these are associated targets.  One of these is not detected in \hi\ 21cm absorption, although the upper limit is quite shallow.  In order to estimate molecular column densities, the excitation temperature was adopted from the original reference when physically justified, or was taken as the default, T$_x$ = 15K, for CO.   Estimating the spin temperature for \hi\ is more problematic.  For intervening absorbers, based on the Milky Way ISM, a value of 100\,K is generally taken.  The spin temperature may be at least as high as 1000\,K in the associated absorbers \citep[e.g.,][]{Maloney1996}.  In Table~\ref{tab:compare}, we adopt T$_s$ = 1000 and 100\,K for associated and intervening absorbers, respectively.
Figure~\ref{fig:compare} shows \hi\ and H$_2$ column densities for both types of systems. 
Clearly, there is no correlation when the bias of higher absorption depth
at greater distances (i.e., at higher $z$)  
is taken into account; emission is more difficult to
detect and only absorption may be detected.
 The $N$(H$_2$)/$N$(\hi) ratio is in the range of 0.03 -- 2.7 for the associated systems; for the intervening systems, it is  0.2 - 1000, with the top three (22 - 1000) corresponding to the lensed systems.   
The scatter decreases if we adopt T$_s$ = 100~K for both the associated and intervening systems. However, a uniform spin temperature for both types of systems cannot be justified simply on the basis of H$_2$/\hi\ ratios.  The associated systems exhibit broader lines, with a median FWHM of $\sim$150\,\kms\ in \hi\ and --- as discussed in Section~\ref{sec:disc-assoc}--- comprise gas from two distinct components: the circumnuclear disk and the galaxy-wide ISM.  On the other hand, for the intervening \hi\ absorption lines are narrower (FWHM$\sim$40\,\kms) and represent multiple clouds with overlapping velocities from the quiescent galaxy disk (see Section~\ref{sec:disc-inter}).  
  
However, several additional redshift-dependent biases ought to be considered for the detectability of absorption: first, contrary to the dust emission, which benefits from a negative K-correction, the AGN synchrotron emission
intrinsically decreases with frequency, and becomes fainter at higher redshift. 
Second, for a given observing frequency, the lines to be observed at higher redshift correspond to higher J-level; these are increasingly difficult to excite in diffuse media, and may be more easily detected through absorption. Also, at higher redshifts, higher spatial resolution is required to resolve the line emission and detect the underlying absorption signal.

\begin{table*}
{\scriptsize
\begin{threeparttable}
\caption{Comparison of 14 known \hi\ 21cm and molecular-absorption-line systems; the first 7 are associated and the remaining ones are intervening. Columns: (1, 2) favored name and systemic optical redshift; (3) redshift based on \hi\ 21cm absorption peak (4, 6) total \hi\ column density,  adopting $T_{\rm s}$ = 1000\,K (100\,K) for associated (intervening) targets, and integrated optical depth; (5) FWHM of the main absorption component;  (7 - 9): molecular column density, FWHM and total integrated optical based on line in column (10); (11): redshift based on the peak of molecular line. (12) and (13): references for \hi\ 21cm and molecular absorption line. In columns (5) and (8), we also give the full velocity extent of the absorption  in parentheses.}
\vspace{-0.4cm}
\begin{center}
\begin{tabular}{lcccccccccccc}
\hline
\hline
 Target  &z$_e$&z$_{21}^*$& N(HI)& FWHM &$\int\tau$dV&N(H$_2$)& FWHM&$\int\tau$dV&Line& $z_{mol}$ &Ref & Ref \\
         & &   & (\cmsq) & (\kms) & (\kms)      & (\cmsq)  & (\kms) &      (\kms) &     &  & (\hi)  & (Mol.) \\
   (1)   & (2) & (3) & (4) & (5) & (6) & (7) & (8) & (9) & (10) & (11) & (12) & (13) \\      
 \hline  
 \\
J0439+0520 &0.2076 & --    & $<$5e22&$<$200     &$<$28.&2.0e21& 126 (210)& 5.4 &CO(1-0) &  0.2077 & H1  &  C1\\
PKS1200+045&1.2243 & 1.2111&  4.6e21& 100 (800) & 2.52 &1.5e20& 87 (225) &3.1  &CO(2-1) &  1.21276& H2  &  C2\\
PKS1245-197&1.275  & 1.2750&  8.3e21& 200 (1000)& 4.54 &3.6e20& 113 (400)&7.2  &CO(2-1) &  1.26605& H2  &  C2\\
4C+12.50    &0.1217 & 0.1217&  6.6e21& 250 (1800)&  3.6 &1.8e22& 250 (600)& 55.0&CO(3-2) &  0.11806& H3   &  C3\\ 
B1504+377  &0.674  &0.67343&  3.8e22& 130 (175) & 21.0 &6.0e20&  75 (100)&10.4 &CO(2-1) &  0.67343& H4   &  C4\\
B1740-517  &0.4423 &0.44129&  5.0e21& 5  (210)  & 2.7  &1.5e20& 37 (120) & 4.8 &CO(2-1) &  0.44185& H5   &  H5\\
Abell 2390 & 0.2304& 0.2310& 2.5e+22& 154 (1100)& 13.7 &8.0e21& 122 (300)& 28.2&CO(1-0) &  0.2312 &H6,H1&  C1\\
\\
 \hline
\\
PMN0134-0931&2.225  &0.76344 &  9.0e20& 107 (200)& 4.93 &2.0e21& 53 (200)&70.3 &CO(2-1) & 0.76391&H7&   C5\\
PKS0201+113&3.639   &3.38714 &  1.8e21& 19 (125) & 0.7  &6.6e20& 57 (400)& 6.2 &CO(4-3) &3.38715&H8&   C2\\
B0218+357  &0.944   &0.68466 &  4.0e20&   43 (75)& 2.19 &4.0e23& 20 (30) &20.3 &C$^{18}$O(2-1)&0.68466& H9&C6,C7\\
G0248+430  &1.313   &0.05151 &  9.7e19&   19 (90)& 0.53 &2.9e19& 16 (30) & 0.25&CO(1-0) &0.05151& H10&   C8\\
PKS1406-076&1.494   &1.27464 &  2.5e20&   11 (50)& 0.51 &1.0e23& 77 (155)& 2.4 &HNC(2-1)&1.26949&H11&   C2\\
PKS1413+135&$\sim$0.3&0.24671&  2.0e21&  18 (180)& 10.86&4.6e20&  2  (5) & 3.64&CO(1-0) &0.24671& H12&   C9\\
PKS1830-211&2.510   & 0.88489&  1.8e21& 180 (400)& 10.1 &4.0e22& 40 (400)&30.  &CO(4-3) &0.88582& H13&   C10\\
\\
\hline
\end{tabular}
 {\hi\ -ref -- H1: \citet{Hogan2014}, H2 \citet{Aditya2018}, H3: \citet{Morganti2004}, H4: \citet{Carilli1997}, H5: \citet{Allison2019},  , H6: \citet{Hernandez2008}, ,   H7: \citet{Carilli1993}, H8: \citet{Kanekar2007},  H9: \citet{Koopmans2005}, H10: \citet{Gupta2018oh},  H11: \citet{Gupta2012},  H12: \citet{Carilli1993, Combes2023}, H13: \citet{Kanekar2003}.
  \item  CO-ref --  C1: \citet{Rose2019all}, C2: present work, C3: \citet{Dasyra2012}, C4: \citet{Wiklind1996b15}, C5: \citet{Wiklind2018}, C6: \citet{Wiklind1995}, C7: \citet{Combes1995},  C8: \citet{Combes2019},   C9: \citet{Wiklind1997},  C10: \citet{Wiklind1998}. }
\begin{tablenotes}[flushleft]
  \item[*] $z_{21}$ is the redshift of peak \hi\ 21cm absorption depth, except for \pks, where
  it is the center of the velocity range over which the absorption is detected.
 \end{tablenotes}
\label{tab:compare}
\end{center}
\end{threeparttable}
}
\end{table*}

\subsection{Nature of the gas detected in intervening systems}
\label{sec:disc-inter}

Because of the low surface filling factor of the molecular gas and the compact size of the AGN continuum emission,  the probability of detecting gas in distant galaxies through emission lines may be much higher than through line absorption. 
Indeed, CO and [C~{\sc ii}] emission from strong metal absorbers or high-$z$ DLAs at $z\sim$0.1 - 4 were recently detected with ALMA and NOEMA \cite[][]{Neeleman2016, Neeleman2019, Kaur2021, Kaur2022}.  Typically, the \hi\ column densities along the background quasar are of the order of 10$^{21}$ cm$^{-2}$, and the galaxies have higher far-infrared luminosities  than the Milky Way. Although the projected separation between a galaxy and a quasar can be relatively high, that is,   10-45\,kpc for [C~{\sc ii}] and as much as 100\,kpc for CO, the velocities of emission and absorption signals are reasonably consistent.   The inferred molecular gas masses are in the range of 10$^9$ to 10$^{10.5}$ M$_\odot$. 
 The intervening emission line detection in our sample have masses of $M$(H$_2$) = $7\times10^{11}$\,M$_\odot$ ($z=3.38715$) and $1.5\times10^{10}$\,M$_\odot$ ($z=0.05151$) (see Table \ref{tab:emassoc}). 
 The CO emission detection rate is 2/14, which is $\sim14^{+19}_{-9}$\%. Among 19 DLAs at $z\sim2$ with CO emission searches reported in the literature, the CO detection rate can be as high as  $56^{+38}_{-24}$\% but only for high-metallicity ([M/H]$>$-0.3) absorbers; otherwise it is merely $\sim11^{+26}_{-9}$\% \citep[][]{Kaur2022}.    
Notably, the $z=3.38715$ detection presented here has the lowest absorption metallicity [M/H] = -1.2 \citep[][]{Kanekar2007} and the highest molecular mass among the DLAs investigated for CO emission to date. However, in relation to other DLAs at comparable redshifts, its metallicity is close to average \citep[e.g.][]{Rafelski2012}.

\subsubsection{Individual sources}
 Now, we turn our attention to the molecular-absorption-line detections. Among the seven intervening systems in Table~\ref{tab:compare}, three are associated with lensing galaxies.  Toward PMN0134-0931, molecular absorption separated by $\sim$200\,\kms\ is seen toward two sight lines separated by 5\,kpc \citep[][]{Wiklind2018}. The spatially unresolved 21cm absorption spectrum shows smoother absorption coinciding with most of the molecular absorption \citep[][]{Kanekar2003}. The \hi\ and molecular absorption line column densities in Table~\ref{tab:compare} correspond to the deeper 21cm absorption component with $N$(CO)/$N$(\hi) = $2.2\times10^{-4}$ ($N$(H$_2$)/$N$(\hi) = 2.2).  The other absorption component exhibits different properties,  with an  abundance ratio that is lower by a factor of $\sim5$.  As previously noted, for PKS0201+113, the 21cm and H$_2$ absorption do not coincide, the peaks are separated by about $\sim$25\,\kms\  \citep[][]{Srianand2012}.    Interestingly, the 21cm absorption and CO absorption components coincide rather well; that is, within 5\,\kms (Fig.~\ref{fig:OQ208}). These may be good candidates for constraining the variation of  fundamental constants of physics using CO and 21cm absorption, albeit with better quality spectra. The total integrated \hi\ column density in Table~\ref{tab:compare} provides $N$(CO)/$N$(\hi) = $3.7\times10^{-5}$ ($N$(H$_2$)/$N$(\hi) = 0.4).  This is a factor of two higher if only the \hi\ in the main absorption component is considered. In general, the CO absorption does not extend over all 21cm components because of the difference in radio continuum extent.  The H$_2$ column density based on Lyman and Werner band absorption lines is $N$(H$_2$) = $10^{14.6 - 16.0}$\,\cmsq\ \citep[][]{Srianand2012}, which is about five orders of magnitude less than the value inferred from the CO absorption.  This and the  disparity between the H$_2$ and 21cm/CO absorption-line shifts suggest parsec-scale structure in the cloud.

 The \hi\ 21cm absorption toward another lensed system, B0218+37, is much broader than the molecular line but the absorption peaks match well within 10\,\kms\ \citep[][]{Wiklind1995}.  The $N$(CO)/$N$(\hi) = 0.1 ($N$(H$_2$)/$N$(\hi) = 10$^3$). The CO and $^{13}$CO lines are optically thick, and the column density is 
 obtained through the C$^{18}$O(2-1) line \citep{Combes1995}. We must caution here that, given the different continuum sizes, the \hi\ and H$_2$ column densities are not averaged over the same apertures.
As previously noted, the atomic and molecular absorption also match well for G0248+430, with $N$(CO)/$N$(\hi) = 2.9 $\times$10$^{-4}$ ($N$(H$_2$)/$N$(\hi) = 2.9), and 90\% of the 21cm optical depth is within the extent of molecular absorption. Toward PKS1413+135, the \hi\ absorption is broad but the molecular absorption is extremely narrow, and $N$(H$_2$)/$N$(\hi) = 0.2. PKS1830-211 is also a lensed system.  In particular, it is known to be rich in gas and dust, exhibiting more than sixty molecular transitions \citep[][]{Muller2014}.  Both the \hi\ and CO absorptions are broad ($\sim$400\,\kms), and the overall $N$(H$_2$)/$N$(\hi) = 22, which is among the highest levels known for this kind of object.

The case of PKS1406-076 is intriguing, with no overlap between the molecular and 21cm line absorptions.  The background radio source is a blazar that exhibits proper motion (0.3\,mas\,yr$^{-1}$) and complex morphology at mas scales over 180\,pc. 
The \hi\ 21cm absorption profile in the GMRT spectra shows three well-detached components and tentative variability over seven months \citep[][]{Gupta2012}.  The disparity between the 21cm and mm molecular absorptions is attributable to the parsec-scale structure in the cloud; that is, the radio and mm sight lines trace different media.   Unfortunately, the presence of \hi\ 21cm absorption corresponding to the HNC absorption is unknown. The GMRT observation to detect \hi\ 21cm absorption used a narrow (1\,MHz) bandwidth to target the \mgii\ absorption and did not cover the 21cm line frequency corresponding to the HNC absorption. 

\subsubsection{Summary}
In summary, in the case of intervening absorbers, the \hi\ absorption is broader and the narrower molecular absorption line is generally close to the \hi\ absorption peak, albeit shifted by 10-20\,\kms.  The exceptions are PKS1406-076 and the three lensed systems where the situation is complicated by multiple sightlines toward the lensed source.   Nevertheless, the observed velocity structure of \hi\ and the molecular absorption profiles imply sight lines piercing through layered molecular clouds; that is, molecular gas embedded in outer layers of predominantly \hi\ gas. 

\subsection{Nature of the gas detected in associated systems}
\label{sec:disc-assoc}

Both associated targets, \pks\ and \pkss, detected in molecular absorption are GPS sources.  In general, such sources are extremely compact ($<1$\,kpc) and are believed to represent the early stages of evolution of powerful radio galaxies \citep[][]{Odea21}. They are known to exhibit the highest \hi\ 21cm absorption detection rates \citep[$\sim$45\%;][]{Gupta2006}. 
Indeed, five out of six CO emission-line detections from our sample are also GPS sources (Table~\ref{tab:emassoc}). In all five cases,
the CO emission line peak coincides with the systemic optical redshift within $\sim$200\,\kms (Fig.~\ref{fig:em8}).   The CO emission line widths (FWHM) are 200 - 800\,\kms, implying that these may have contributions from the gas clouds from {\it (i)} the host galaxy ISM through which the young radio source is expanding, and {\it (ii)} the circumnuclear disk, especially the regions close to the central AGN.
 The emission and absorption detected toward Abell\,2390 is complex: the CO emission is extended and exhibits large-scale outflows and the \hi\ 21cm absorption has a narrow feature with broader components on either side or covering the entire velocity \citep[][]{Hernandez2008, Rose2019all}.
Nevertheless, the \hi\ absorption line peak in all but one source (i.e., \pks) is also within the velocity range over which CO emission is detected.  Therefore, to first-order, \hi\ 21cm absorption and CO mm emission lines are originating from the same regions.

In general, at cm wavelengths, the absorption is more likely to be detected in front of radio jets and lobes, whereas at mm wavelengths the gas in front of the flat spectrum ``core''  dominates the detected signal.  In addition, molecular gas is embedded in ISM clouds with outer layers of higher atomic gas fraction. Both these factors, that is, geometrical effects ---due to orientation and the frequency-dependent structure of the radio source--- and varying molecular gas fraction within the cloud may lead to differences in the observed properties of the gas detected in atomic and molecular absorption lines.

\begin{figure}
\begin{center}
\includegraphics[clip,width=0.5\textwidth,angle=0]{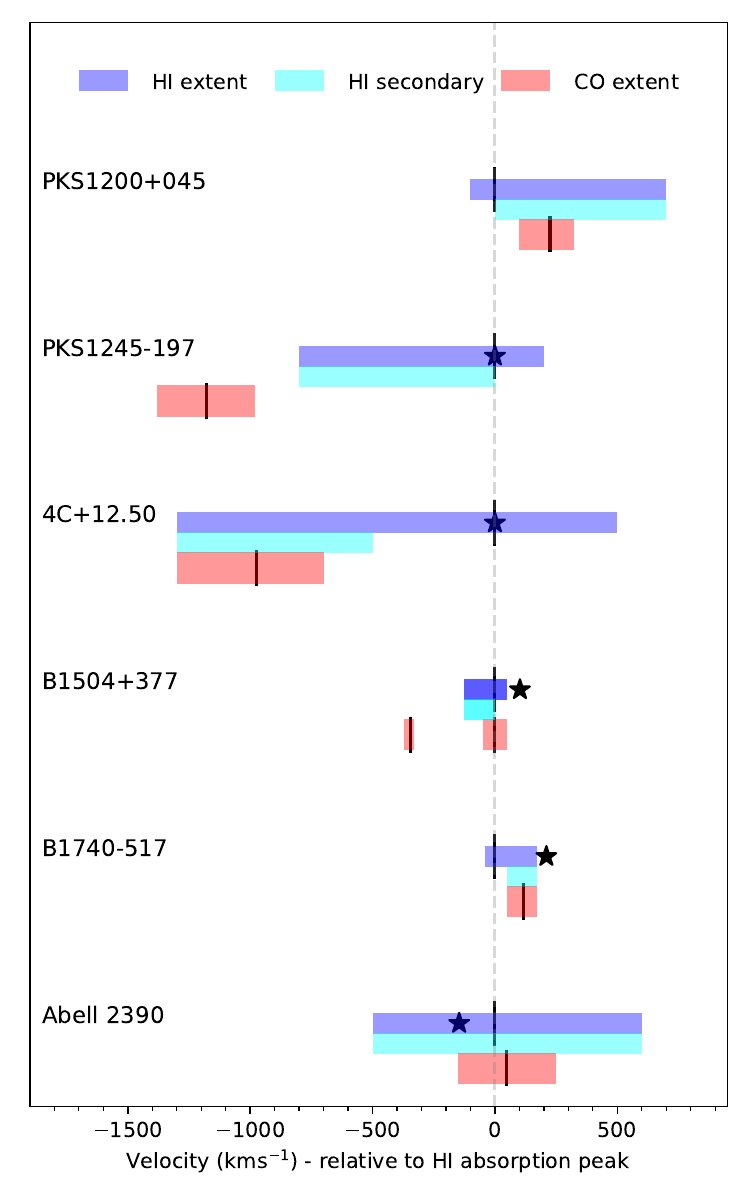}
\vskip+0.0cm
\caption{Velocity offsets of \hi\ 21cm and CO absorption lines.  The zero of the velocity scale corresponds to $z_{21}$.  For \hi, both the extent of the full and secondary absorption are shown. The vertical ticks mark the location of \hi\ and CO absorption peaks, and the systemic velocity ---except for \pks\,--- based on optical emission lines is shown with a star $\star$.  
} 
\label{fig:assocvel} 
\end{center}
\end{figure}

In addition to the two associated CO absorbers reported here, associated molecular absorption lines have been detected in five other systems (see Table~\ref{tab:compare}). In general, associated \hi\ 21cm absorption lines are known to exhibit a deep narrow component and a broad shallow component.  The former, when coincident with the systemic redshift, is believed to originate from the circumnuclear disk or torus.  The latter, depending on whether it is blueshifted or redshifted with respect to the systemic redshift, is interpreted as outflowing gas due to jet--ISM interaction or infalling gas that may eventually fuel the AGN, respectively. In Fig.~\ref{fig:assocvel}, we present the extent of the overall \hi\ 21cm absorption (in blue) and the associated secondary (i.e., broader/shallower) component (in cyan) for six AGNs\footnote{BCG J0439+0520 with no \hi\ 21cm absorption is excluded.}. 
In five of the six cases, the \hi\ absorption peak, shown as a vertical blue line, matches the systemic redshift within $\pm$200\,\kms. 
Interestingly, there is a tendency among the associated systems to present CO absorption (in red) coincident with the weaker 21cm absorption component.

\subsubsection{Individual sources}
For \pks, the observed offset ($\sim$2000\,\kms) between the \hi\ 21cm and CO absorption lines with respect to molecular and optical/UV emission lines is particularly intriguing.  The large velocity offset with respect to the CO emission lines and systemic redshift, and the widths of \hi\ 21cm ($\sim$800\,\kms) and CO ($\sim$200\,\kms) absorption lines suggest that   the absorbing gas represents a distinct and scarce population of cold gas clouds in the host galaxy. The associated radio emission at 1.4\,GHz with a projected linear size of 60\,mas (500\,pc at $z_{abs}$) exhibits three components, one of which lies at one end and contains more than 90\% of the total flux density \citep[][]{Liu2007}.  The nature of this one-sided morphology as a core--jet structure is confirmed by the higher-frequency milliarcsecond-scale images at 2.3, 4.3, and 8.7\,GHz available from the Very Long Baseline Array (VLBA) calibrator catalog\footnote{https://obs.vlba.nrao.edu/cst/}.  The one-sided core--jet morphology and orientation of the jet axis likely ensures that the sight lines to the radio source do not pass through the circumnuclear disk detected in CO emission. The absorbing gas may then correspond to quiescent clouds in the host galaxy ISM or gas interacting with the jet.  The large velocity widths of \hi\ 21cm ($\sim$800\,\kms) and CO ($\sim$200\,\kms) absorption lines suggest that the latter is true,  with the gas being  detectable in absorption primarily due to the orientation of the radio source.

For \pks\ one possibility is that only the radio ``core'' is responsible for the absorption detected at cm and mm wavelengths. The \hi\ absorption consists of two components separated by $\sim$200\,\kms\ and CO absorption is coincident only with the secondary weaker component (Fig.~\ref{fig:assocvel}).  Assuming complete coverage of the radio core
for the total integrated columns of the atomic and molecular gas, and $T_{\rm s}$ = 1000\,K, we estimate the CO-to-\hi\ column-density ratio, $N$(CO)/$N$(\hi) = $3\times10^{-6}$ , which means $N(H_2)$/$N$(\hi) =  $3\times10^{-2}$.
Alternatively, it is also possible that only the CO and secondary \hi\ absorption components are originating from the gas in front of the radio core.  The molecular to \hi\ column density ratio would then be a factor of two higher. The deeper \hi\ absorption component would then correspond to the gas in front of the weaker radio components associated with the jet and the lobe.  The projected distance of this gas component would be at least a few hundred parsecs from the core, which would explain the absence of CO absorption at the corresponding velocities.  

The second possibility for \pks\ is similar to the scenario observed for 4C+12.50: a CSS source hosted in a gas-rich Seyfert galaxy.  At 1.4\,GHz, 4C+12.50 exhibits a distorted radio morphology with an overall extent of 150\,mas ($\sim$325\,pc).  The mas-scale spectroscopy reveals the deep, narrow \hi\ 21cm absorption component that is  detected toward the fainter northern lobe rather than the core or the brighter southern hotspot \citep[][]{Morganti2004}. The weaker \hi\ absorption component coincides with the CO absorption. The CO-to-\hi\ column density ratio is $6\times10^{-4}$ ($N(H_2)$/$N$(\hi) = 6) for $N$(\hi) = $3\times10^{21}$\,\cmsq\ based on the Na~{\sc i} or \hi\ 21cm absorption ($T_{\rm s}$ = 1000\,K) associated with the absorption component coincident with the molecular absorption.  This is about 200 times higher than the ratio observed for \pks\ and is likely due to the fact that 4C+12.50 {\it (i)} is hosted in a gas-rich Seyfert galaxy, and {\it (ii),} being a radio galaxy, PKS\,1200$+$045 is oriented such that the sight line to the radio core passes through the circumnuclear disk.

In contrast to \pks, the CO absorption associated with \pkss\ is offset with respect to the 21cm absorption by $-$1180\,\kms\  but is within the range over which CO emission is detected (Fig.~\ref{fig:PKS1245}).  The 21cm absorption coincides well with the CO emission (Fig.~\ref{fig:PKS1245}).  At 2.3 and 8.6\,GHz, the radio source exhibits two prominent, steep spectrum ($\alpha<$-0.5) components of similar strength, separated by $\sim$25\,mas \citep[][]{Sokolovsky2011}.  The radio core is not detected in these low-frequency images but may be responsible for the CO absorption.  The sight lines at cm wavelengths, especially those toward the counter jet and lobe, pass through the circumnuclear disk as well as the host galaxy ISM.  The nondetection of \hi\ 21cm absorption ---adopting $T_{\rm s} = 1000$\,K and a line FWHM = 200\,\kms\ based on the detected CO absorption--- corresponds to a CO-to-\hi\ column density ratio of $>7\times10^{-5}$ ($N(H_2)$/$N$(\hi) $>$ 0.7). This is significantly higher and lower than the ratios observed for \pks\ and 4C+12.50, respectively.   Overall, both the 21cm and CO absorption lines seem to be tracing the galaxy-wide cold gas from the ISM and the circumnuclear disk detected in CO emission. The disparity between \hi\ and CO absorption lines is most likely due to the differences in the sight lines at cm and mm wavelengths.

Molecular absorption lines have also been detected toward B1504+377, which is hosted by a disk galaxy. The associated radio emission at cm wavelengths consists of a dominant `core' component, with some diffuse emission at a distance of 55\,mas (390\,pc at $z$=0.674).  There is coincidence between the broader \hi\ and CO absorption (FWHM$\sim$75\,\kms) components  at $z$ = 0.67343, both of which are coincident with the systemic velocity \citep[][]{Wiklind1996b15, Carilli1997}. However, no \hi\ absorption is detected in the narrow (FWHM$\sim$15\,\kms) molecular absorption component at $z=0.67150$ (see Fig.~\ref{fig:assocvel}).
For the broader component, the CO-to-\hi\ column-density ratio is $2\times10^{-6}$ ($N(H_2)$/$N$(\hi) $>$ 0.02).  The orientation and the morphology of the radio source suggest that it is tracing the  gas through the circumnuclear disk as intercepted by the core.  The nondetection of \hi\ 21cm absorption in the narrow CO absorption component corresponds to a column density ratio of $>3\times10^{-5}$ ($N(H_2)$/$N$(\hi) $>$ 0.3),  which is comparable to other similar systems discussed here but is at least a factor of ten higher than the broader component.  Based on the large shift ($-$450\,\kms) but narrow width (15\,\kms), it likely represents a cloud in the inner circumnuclear disk.  The \hi\ 21cm absorption nondetection is due to the relatively high CO-to-\hi\ abundance ratio.

Another molecular absorber is B1740-517, which is a GPS source associated with a radio galaxy (Sy).  At 2.3\,GHz, radio emission is resolved into two components separated by $\sim$50\,mas ($\sim$300\,pc at $z$=0.442) and the CO(2-1) absorption coincides with a weaker \hi\ absorption component \citep[][]{Allison2019}.  No CO absorption is detected at the narrow, deep 21cm absorption peak, which is blueshifted with respect to the CO absorption and the systemic redshift by $\sim$130\,\kms. It is unclear whether the two radio components are core--jet or two lobes.  The CO-to-\hi\ column density ratio for the absorption component with \hi\ and CO absorption, $2\times10^{-5}$ ($N(H_2)$/$N$(\hi) $=$ 0.2),  is comparable to those of other associated systems. It may represent gas closer to the central AGN, whereas the narrow component 
represents distant gas toward the jet and the lobe.
\subsubsection{Summary}
Overall, CO absorption is generally detected in broader and weaker 21cm absorption components with typical CO-to-\hi\ abundance ratios of  $10^{-6}$ -- $10^{-5}$ ($N(H_2)$/$N$(\hi) = 0.01 - 0.1).  Except for B1504+377, the CO absorption never coincides with the deep, strong 21cm absorption component.   This suggests that absorbing gas has two different phases: one phase near galaxy centers with a larger CO-to-\hi\ abundance ratio, and another with lower CO abundance that shows only \hi\, and might correspond to the gas in outer regions of galaxies. These two phases are well known through emission-line observations of nearby galaxies \citep[e.g.,][]{Bigiel2012}.  The detection rate  of molecular absorption from our survey is 2/16, which equates to $13^{+16}_{-8}$\%. 
As mentioned above, the detection of molecular absorption is not expected in all associated systems with 21cm absorption because of the smaller filling factor of the molecular component, and the much smaller continuum sizes
at mm frequency.

\section{Conclusions}
\label{conclu}

We searched for molecular emission and absorption in 30 \hi\ 21cm absorbers at $0.1<z<4$. Of these, 16 are associated with the AGN and the remaining 14 are from intervening galaxies. We report the detection of CO emission in 8 systems, of which 5 are new.  The derived molecular masses, assuming standard conversion ratios, range from 10$^9$ to 7$\times$ 10$^{11}$ M$_\odot$.  The majority of these correspond to associated systems belonging to young radio AGN, that is, intrinsically compact radio sources with a peaked radio spectrum. Four of them are at low redshift ($z<$0.25). Intervening galaxies that show CO emission are rarer (2/14 = $\sim14^{+19}_{-9}$\%). 
The two detections reported here from our sample have masses of $M$(H$_2$) = $7\times10^{11}$\,M$_\odot$ (PKS0201+113; $z=3.38715$) and $1.5\times10^{10}$\,M$_\odot$ (Q0248+430; $z=0.05151$).  Notably, the $z=3.38715$ detection presented here has the lowest absorption metallicity ([M/H] = -1.2) and the highest molecular mass among the DLAs investigated for CO emission to date \citep[][]{Kaur2022}. We note that the average metallicity of DLAs at such high redshifts is [M/H] $\sim$ -1.5 \citep[][]{Rafelski2012}.
We also stacked together the spectra from 21 undetected sources (10 associated and 11 intervening) and find a hint of emission, that is, at a flux level of about an order of magnitude lower than the individual detections. It is not possible to draw conclusions on the corresponding masses because of the varied distances.

Millimeter molecular absorptions are still very rare at moderate and high redshift: only ten systems have been found \citep{Combes2008, Dasyra2012, Wiklind2018, Allison2019, Combes2019, Rose2019all}: five are associated absorbing systems, and the remaining five are intervening systems, of which three are gravitational lenses. Here, we report four systems showing absorption line detections, two associated and two intervening.   The two associated systems, \pks\ ($z$ = 1.2111) and \pkss\ ($z$ = 1.2750), were detected in high-spatial-resolution NOEMA follow-up observations of our IRAM 30m detection of CO emission, and demonstrate the complexity faced in detecting absorption in the presence of emission. Alternatively,  it is possible to detect absorption in the high-density tracers, such as HCO$^+$, HCN, or HNC, which are not detectable in emission. This is indeed the case for the intervening system toward PKS1406-076 ($z$ = 1.27464) reported here. 

The other two intervening detections are toward PKS0201+113 ($z$ = 3.38714) and Q0248+430 (0.05151) \citep{Combes2019}. For PKS0201+113, the 21cm and H$_2$ (ultraviolet) absorption components do not coincide; the peaks are separated by about $\sim$25\,\kms\  \citep[][]{Srianand2012}. Interestingly, the 21cm absorption and CO absorption components coincide rather well (i.e., within 5\,\kms; Fig.~\ref{fig:OQ208}), and the latter could be good candidates with which to constrain the variations of fundamental constants of physics.  The disparity between H$_2$ and 21cm/CO absorption line shifts suggests parsec-scale structure in the cloud. The case of PKS1406-076 is even more intriguing, with no overlap between the molecular and 21cm line absorptions.  The disparity between the 21cm and mm molecular absorptions is attributable to complex radio morphology; that is, radio and mm sight lines tracing different media.

 We compare the atomic and molecular column densities of 14 known high-redshift ($z>0.1$) molecular absorption line systems. Among associated systems, CO absorption is generally detected in the broader and weaker 21cm absorption component, with a typical CO-to-\hi\ abundance ratio of  $10^{-6}$ -- $10^{-5}$ ($N(H_2)$/$N$(\hi) = 0.01 - 0.1).  The CO absorption very rarely coincides with the deep and narrow 21cm absorption component.   This suggests that absorbing gas has two different phases: one phase near to galaxy centers with a larger CO-to-\hi\ abundance ratio, and another with lower CO abundance that shows only \hi\, and might correspond to the gas in outer regions of galaxies. These two phases are well known through emission line observations of nearby galaxies \citep[e.g.,][]{Bigiel2012}.       

 In the case of intervening absorbers, the \hi\ absorption is broader and the narrower molecular absorption line is generally close to the \hi\ absorption peak, although shifted by 10-20\,\kms.  The exceptions are PKS1406-076 and the three lensed systems where the situation is complicated by multiple sightlines toward the lensed source. Nevertheless, the observed velocity structure of \hi\ and molecular absorption profiles imply sight lines piercing through layered molecular clouds; that is, molecular gas embedded in outer layers of predominantly \hi\ gas. 

 With the advent of large surveys at cm wavelengths, the number of \hi\ 21cm absorbers is expected to steeply increase.  However, currently, \hi\ 21cm line spectroscopy at z$>$0.5 can be performed only at spatial resolutions of $>5^{\prime\prime}$, which may be insufficient to reveal the origin of the absorbing gas.  The work presented here demonstrates that this limitation can be overcome through the combination of mm emission and absorption spectroscopy, and helps to elucidate the role played by cold gas in the evolution of normal and active galaxies.

\begin{acknowledgements}
We sincerely thank the referee for very useful and constructive comments, 
and a thorough reading of the manuscript. 
This work is based on observations carried out with the IRAM-30m telescope, and the NOEMA Interferometer. IRAM is supported by INSU/CNRS (France), MPG (Germany), and IGN (Spain). We acknowledge the help of the IRAM teams for the observations and thank Vinodiran Arumugam for his support with the data reduction. The data were processed using the Gildas package.
This project has benefited from support from the Programme National Cosmologie et Galaxies.
We made use of the NASA/IPAC Extragalactic Database (NED), and of the HyperLeda database (http://leda.univ-lyon1.fr).
\end{acknowledgements}

\bibliographystyle{aa}
\bibliography{PKS12.bib}

\end{document}